\documentclass[
 reprint,
superscriptaddress,
 amsmath,amssymb,
 aps, showkeys,
prb,
]{revtex4-2}

\usepackage{graphicx}
\usepackage{dcolumn}
\usepackage{bm}

\begin{document}

\title{Non-Hermitian Exchange as the Origin of Chirality-Induced Spin Selectivity}

\author{Pius M. Theiler}

\email{pius.theiler@nrel.gov}
\affiliation{
National Renewable Energy Laboratory,
Golden, CO, USA
}

\author{Sander Driessen}
\affiliation{
Department of Chemistry, Duke University
, Durham NC,  USA
}
\affiliation{Department of Applied Physics, Eindhoven University of Technology, Eindhoven 5612 AZ, The Netherlands}

\author{Matthew C. Beard}

\affiliation{
National Renewable Energy Laboratory,
Golden, CO, USA
}

\date{\today}

\begin{abstract}
For over two decades, the role of structural chirality in spin polarization has been widely investigated, with implications for the origins of life, catalysis, and quantum phenomena. Yet, it remains unclear whether all chirality-induced spin selectivity (CISS) effects arise from a common mechanism. We show that breaking all mirror symmetries in structurally chiral electron systems enforces a twin-pair electron exchange, inherently violating both parity ($\mathcal{P}$) and time-reversal ($\mathcal{T}$) symmetry while preserving combined $\mathcal{PT}$ symmetry of the Hamiltonian. This exchange produces chiral quantum states where electron spin and motion are intrinsically linked, a key feature of CISS. At interfaces, these states drive spin and charge accumulation via spin-momentum locking. Our findings establish a new paradigm connecting quantum statistics, non-Hermitian physics, and spin transport with structural chirality. This framework unifies all observed CISS effects and provides guiding principles for designing chiral materials for spintronic and quantum applications.

\end{abstract}
\maketitle

\section{\label{sec:Intro}Introduction}
Phenomena encompassing the interaction between the electron's spin and structural chirality where either the opposite enantiomer or opposite spin, lead to different observables, whereas flipping both results in no observable change is dubbed the chirality-induced spin selectivity (CISS) effect\cite{Bloom2024}. Exploiting how electrons in chiral structures behave will have applications in diverse fields such as quantum physics, chemistry,  biology, medicine, and material science. Chiral structures and their interactions with electrons may have played an important role in the origin of life\cite{Bloom2024}. To unlock this potential of CISS, a detailed mechanistic understanding is needed.

\begin{figure*}
\includegraphics[width = 0.95\textwidth]{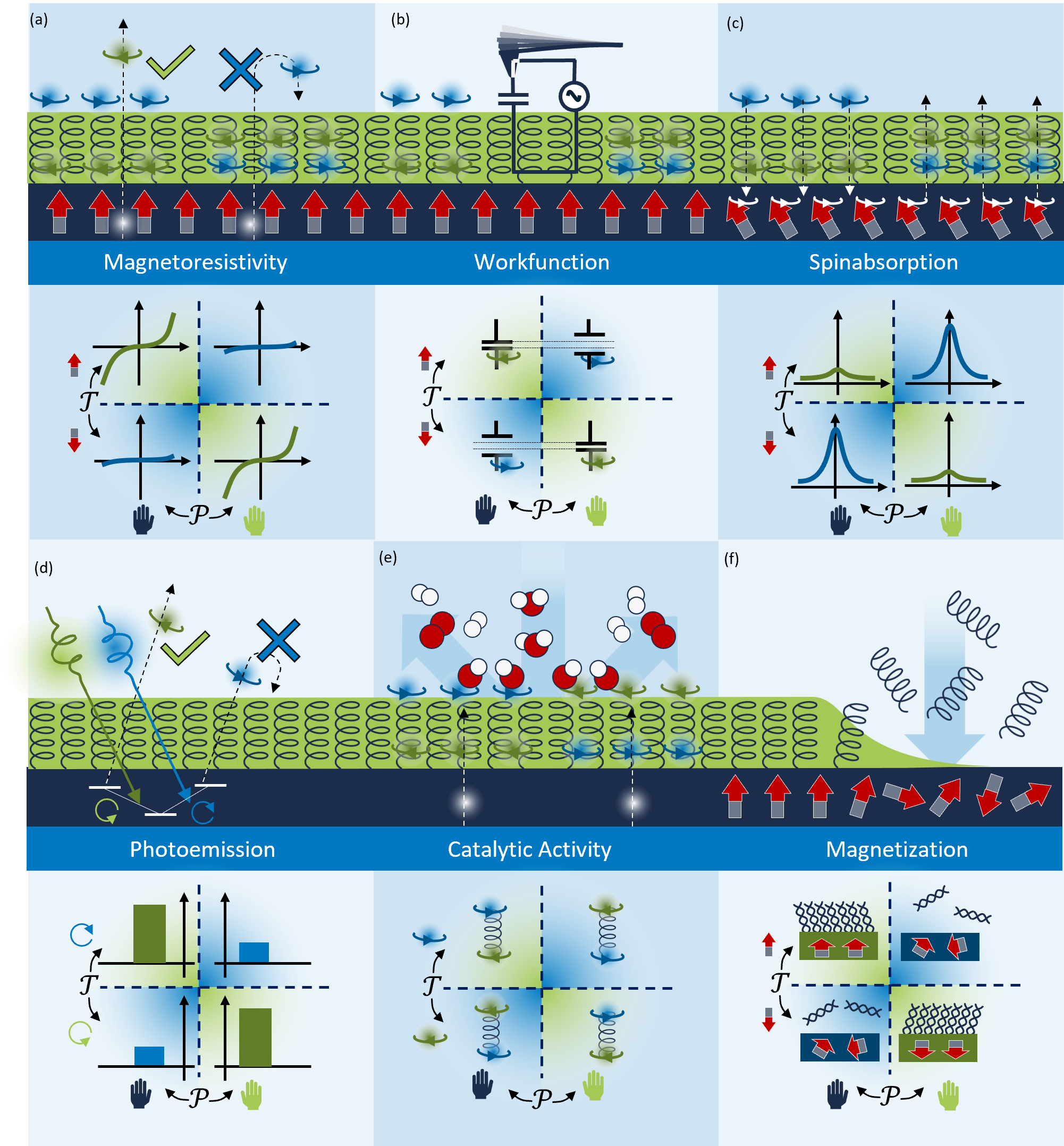}
\caption{ Different classes of CISS experiments and their symmetries. Magneto-resistance experiments (a) show to distinct I-V curves depending upon the enantiomer and magnetization of the ferromagnetic substrate, Measurement of a quantum capacitance and contact potential difference using Kelvin probe force microscopy (b) shows a different charge and spin distribution for opposite enantiomer, spin pumping experiments(c) show different charge currents or absorbance characteristics upon switching magnetization or enantiomers.  The photoelectric effect on chiral materials lead to the ejection of preferred spins (d) where the ejection probability is different for changing either enantiomer or circular polarization, but  remains unchanged flipping both. Enantiomer induced spin-selective catalysis (e) provide spin polarized charge currents. The adsorption of enantiomer (f) leads to a permanent magnetization of the substrate. All classes of experiments are invariant under $\mathcal{PT}$ operation of the experimental parameters.}
    \label{fig:Figure1}
\end{figure*}

Despite tremendous experimental and theoretical efforts, the origin of CISS remains unclear~\cite{Aiello2022,Bloom2024,Evers2022}. Previous theoretical attempts have been limited to specific experiments, but have proven difficult to generalize. Figure \ref{fig:Figure1} summarizes the six most common experimental observation classes attributed to CISS. The central question addressed here is how to describe CISS consistently in magneto-resistance\cite{Al-Bustami2022}(Fig.\ref{fig:Figure1}a), quantum capacitance\cite{Theiler2023} (Fig.\ref{fig:Figure1}b), chiral spin-to-charge conversion\cite{Moharana2025,Inui2020,Sun2024} (Fig.\ref{fig:Figure1}c), photoemission\cite{Abendroth2019} (Fig.\ref{fig:Figure1}d), chirality dependent catalytic activity\cite{Liang2022} and spin-sensitive reactions\cite{Bhowmick2022} (Fig.\ref{fig:Figure1}e), magnetization upon chiral molecular absorption\cite{BenDor2017} and enantiomer selective absorption\cite{Koyel2018} (Fig.\ref{fig:Figure1}f) in a unified way. These experimental observations\cite{Al-Bustami2022,Theiler2023} indicate that there is a simultaneous breaking of both time-reversal, $\mathcal{T}$ and parity, $\mathcal{P}$ symmetries, but the product, $\mathcal{PT}$ remains invariant. 

 Breaking of $\mathcal{P}$  symmetry occurs by definition in a chiral structure. Usually, breaking of $\mathcal{T}$  symmetry in CISS experiments is argued to occur through dissipative processes\cite{Evers2022,Bloom2024}. However, this does not align with observations that (a) show long lived equilibrium processes, e.g. persistence of magnetization upon interacting with chiral molecules\cite{BenDor2017} and quantum capacitance\cite{Theiler2023}, or (b) show chirality dependent magnetoresistance at zero bias in a two-terminal spin-valve \cite{Tirion2024,Al-Bustami2022}. In addition, charge-to-spin conversion usually requires large spin-orbit coupling (SOC) but this is in constrast to many observations of CISS in organic molecules with little to no SOC.  

Our proposed mechanism of $\mathcal{T}$symmetry breaking does not require dissipation and is inherent to chiral systems.  We propose that chiral systems require by symmetry the exchange interactions of at least four electrons. We show that such an exchange interaction gives rise to a term in the Hamiltonian of the form $i \hat{\sigma}\cdot\hat{p}$ , where $\hat{\sigma}$ is the spin angular momentum and $\hat{p}$ is the linear momentum operator. This exchange term results in a non-Hermitian Hamiltonian that breaks $\mathcal{T}$ symmetry but conserves both energy and angular momentum in a closed system. In finite systems a Non-Hermitian skin effect\cite{Xiong2018,Kunst2018} results where opposite spin orientations are exponentially localized on the opposing boundaries. In this view, CISS observables are caused by \textit{both} - an interface that localize the spin states on opposing sides and the bulk chiral symmetry dictates the exchange interaction energy, but notably \textit{does not} rely upon SOC or dissipation. This new framework paves the way to settle long-standing questions regarding CISS, which we outline here.

\section{About spin-statistics and structural chiral symmetry}
The spin-statistics theorem governs basic properties of condensed matter systems: exchanging two indistinguishable fermions introduces a global phase shift of $\pi$ in the wavefunction~\cite{Streater1964}, underpinning phenomena such as the Pauli exclusion principle and magnetic ordering~\cite{Dresselhaus2008}. 

This condition arises from the symmetry properties of the many-particle wavefunction \(\psi(x_1, x_2, \dots, x_n)\) under particle exchange, where each particle variable \(x_j = (\vec{r}_j, \hat{\sigma}_j)\) includes both spatial coordinates and spin. For indistinguishable particles, the wavefunction must transform consistently under exchange, regardless of whether it factorizes into separate spin and spatial parts. In particular, exchanging particles \(1\) and \(2\) yields
\begin{align}
   \psi(x_1, x_2, \dots, x_n) =& r\,\psi(x_2, x_1, \dots, x_n)   \\
   =& r^2\,\psi(x_1, x_2, \dots, x_n),
   \label{Eq.c=1}
\end{align}
which implies \(r = \pm 1\). The case \(r = +1\) corresponds to bosonic, symmetric wavefunctions,
and \(r = -1\) corresponds to fermionic, antisymmetric wavefunctions.

Applying this framework to structurally chiral multi-electron systems reveals a fundamental contradiction. In a system of two pointlike particles, operations such as particle exchange, parity operation, time reversal, and spatial inversion are physically indistinguishable - they all map the system onto an equivalent configuration. This breaks down in chiral multi-electron systems. The absence of mirror symmetry makes a pairwise particle exchange physically distinguishable: it flips the chirality of the system. This violates the foundational assumption that exchanged configurations are observationally indistinct to an external observer.  Notably, this inconsistency can be resolved if two correlated pair-exchanges occur simultaneously, referred to here as a twin-exchange, preserving the global chirality of the system.

\begin{figure}
\includegraphics[width = 0.45\textwidth]{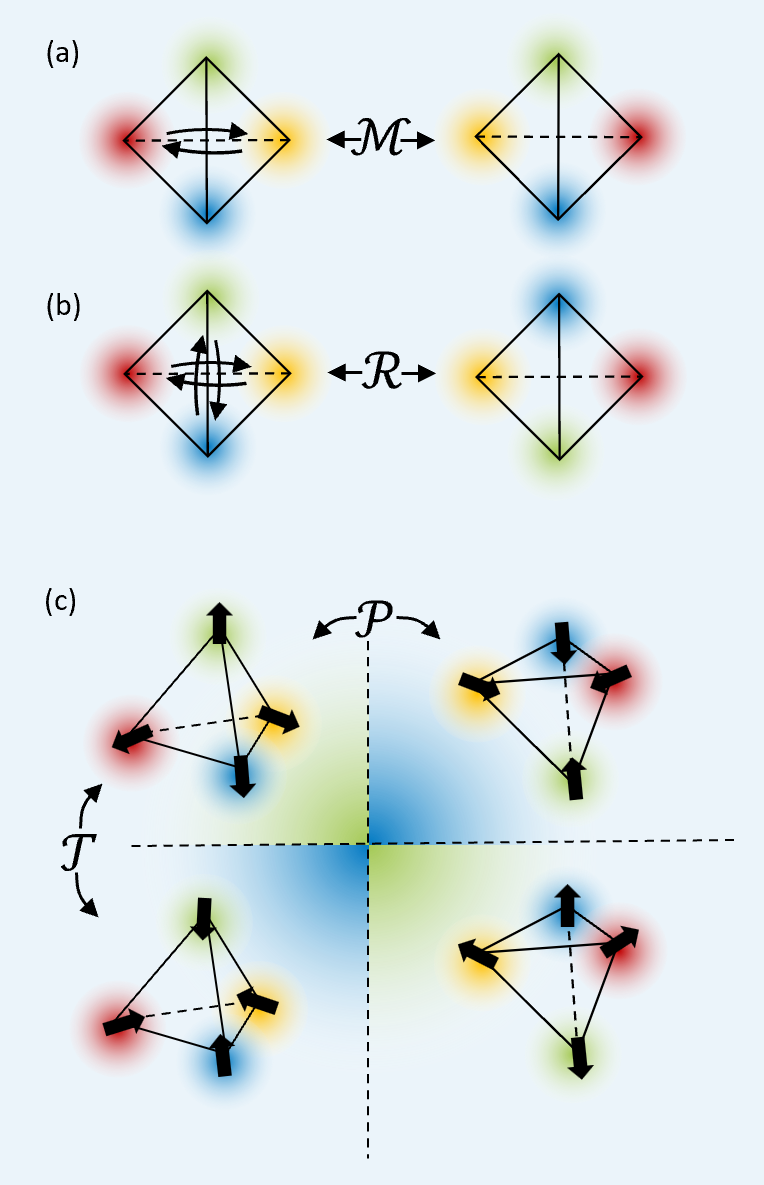}
\caption{  Chiral tetrahedron illustrating four distinct, color-coded multi-electron wavefunctions $\Psi^\xi$ with encoding non-local chirality. The vertex shading indicates the total charge density of the multi-electron wavefunction $\Psi^\xi$, while spinors represent the spin texture. (a) a single pair exchange of $\Psi^\xi$ acts as an effective mirror transformation $\mathcal{M}$, equivalent to a combined parity transformation and rotation.
(b) two correlated pair-exchanges rotates $\mathcal{R}$ while preserving its chirality.  
(c) figure illustrates how structural chirality leads to the breaking of both parity $\mathcal{P}$ and time-reversal $\mathcal{T}$ considering charge- and spin-densities of the multi-electron wavefunctions $\Psi^\xi$.}
    \label{fig:Figure2}
\end{figure}

To illustrate this point, Fig.~\ref{fig:Figure2}a and b show a multi-electron chiral distribution where the vertex shading represents the electron density of the entire many-body wavefunction. This emergent state reflects the collective properties rather than single-particle behavior. Charge and spin densities transform under different symmetry constraints\cite{Liu2022}: while the charge distribution follows the conventional spatial point group symmetry, the spin texture transforms under the spin point group symmetry\cite{Dresselhaus2008}, enabling non-trivial spin arrangements even in the absence of spin-orbit coupling\cite{Liu2022}. A single exchange flips the enantiomer, e.g., in Fig.~\ref{fig:Figure2}a a single exchange is equivalent to a mirror operation, $\mathcal{M}$ , whereas a twin exchange induces a chirality-preserving $\pi$ rotation, e.g. Fig.~\ref{fig:Figure2}b . This concept extends naturally to molecules with multiple stereogenic centers~\cite{Capozziello2003} or chiral axes. 

Formally, identifying chirality requires comparing spatial relationships between multiple particles—an inherently nonlocal operation\cite{Kusunose2024}. Relativistic causality prohibits local observers from instantaneously accessing the full geometric arrangement of the system.   To circumvent this limitation, an internal degree of freedom, $\xi$ can encode the global chiral information. This internal label is not directly observable via local measurements but plays a crucial role in the transformation properties of the full wavefunction under permutations. 

Inspired by the formalism introduced by Wang and Hazzard~\cite{Wang2025}, the exchange of two adjacent electrons in such a wavefunction can be expressed as

\begin{equation}
\Psi^\xi \left(\{x_i\}_{i=1}^{n}\right)\big|_{x_j \leftrightarrow x_{j+1}} 
= \sum_J (R_j)^\xi_{\;J} \, \Psi^J(\{x_i\}_{i=1}^{n}),
\end{equation}

where the superscript $\xi$ labels internal symmetry states, and the matrices $R_j$ encode how the exchange operation on particle $x_j \leftrightarrow x_{j+1}$ acts within this internal Hilbert space. The four different $\Psi^\xi$ are illustrated as color coded vertex clouds in Fig.\ref{fig:Figure2}. The index $J$ runs over the full basis of this space. To be consistent with the constraint $r^2 = \pm 1$ in Eq.~\ref{Eq.c=1}, these matrices must satisfy

\begin{equation}
R_j^2 = I, \quad \text{and} \quad R_j R_i = R_i R_j.
\end{equation}

In a chiral system the allowed $R$-matrices representing such exchanges must be carefully constrained. Specifically, only those permutations that preserve the chiral structure and do not induce physically distinguishable effects on the wavefunction are permitted. To construct the allowed $R_j$ matrices,  we can analyze the symmetries of a chiral tetrahedral configuration (Fig. \ref{fig:Figure2}a and b).  The chiral tetrahedron is the simplest structure without mirror symmetry; fewer than four point-like particles always define a mirror plane. In the chiral tetrahedron  the $R$-matrices are restricted to a proper subgroup of the full permutation group—namely, those that square to the identity and commute with each other—ensuring consistency with both indistinguishability and the chiral nature of the system. The full set of vertex permutations forms the symmetric group $S_4$, containing $4! = 24$ elements. However, only the 12 even permutations that preserve the handedness of the tetrahedron form the alternating group $A_4$~\cite{Capozziello2003}. Among these, four elements represent pure $\pi$-rotations (i.e., exchanges that square to identity), and these generate a subgroup isomorphic to the Klein four-group, $V_4$. The remaining elements of $A_4$ include $3$-cycles, which correspond to $2\pi/3$ rotations and thus violate $R_j^2 = I$. Thus, only the Klein-four $V_4$ subgroup remains consistent with the exchange constraint.

\begin{align*}
R_1 =& \begin{bmatrix}
1 & 0 &  0 & 0\\
0 & 1 &  0 & 0\\
0 & 0 &  1 & 0\\
0 & 0 &  0 & 1\\
\end{bmatrix}  R_2 = \begin{bmatrix}
0 & 0 &  0 & 1\\
0 & 0 &  1 & 0\\
0 & 1 &  0 & 0\\
1 & 0 &  0 & 0\\
\end{bmatrix} \\ R_3 =&\begin{bmatrix}
0 & 1 &  0 & 0\\
1 & 0 &  0 & 0\\
0 & 0 &  0 & 1\\
0 & 0 &  1 & 0\\
\end{bmatrix} R_4 = \begin{bmatrix}
0 & 0 &  1 & 0\\
0 & 0 &  0 & 1\\
1& 0 &  0 & 0\\
0 & 1 &  0 & 0\\
\end{bmatrix}
\end{align*}
One can easily see that this residual symmetry group captures the essence of chirality-preserving, spin-rotating twin-exchanges in the tetrahedral configuration. For the non-chiral tetrahedron, applying the same logic gives the elemental permutations

\begin{align*}
R_1' =& \begin{bmatrix}
1 & 0 &  0 & 0\\
0 & 1 &  0 & 0\\
0 & 0 &  1 & 0\\
0 & 0 &  0 & 1\\
\end{bmatrix}  R_2' = \begin{bmatrix}
0 & 1 &  0 & 0\\
1 & 0 &  0 & 0\\
0 & 0 &  1 & 0\\
0 & 0 &  0 & 1\\
\end{bmatrix} \\ R_3' =& \begin{bmatrix}
1& 0 &  0 & 0\\
0 & 0 &  1 & 0\\
0 & 1 &  0 & 0\\
0 & 0 &  0& 1\\
\end{bmatrix} R_4' = \begin{bmatrix}
1 & 0 &  0 & 0\\
0 & 1 &  0 & 0\\
0& 0 &  0 & 1\\
0 & 0 &  1 & 0\\
\end{bmatrix}
\end{align*}

that are covering the full symmetric group $S_4$.

This analysis demonstrates that exchange statistics, dictated by symmetry, differ between chiral and non-chiral systems. In the symmetric, non-chiral case, electron pairs exchange independently captured via the $R'_j$ matrices.  While in chiral systems when one pair exchanges, the other must simultaneously exchange and this is captured via $R_j$ matrices.

For a mirror-symmetric, non-chiral multi-electron system, pair-exchange statistics can be mapped via exchange integrals onto a Hamiltonian of the form  
\begin{equation}
H_{ij} = J(1+\hat{\sigma}_i\hat{\sigma}_j),
\label{Eq:h1}
\end{equation}  
corresponding to the well-known quantum Heisenberg interaction \cite{Hoffmann2020}, where $\hat{\sigma}_{i}$ are Pauli spin matrices. In the Dirac theory—the simplest framework unifying spin and relativity \cite{Dirac1928} -  the analogous role of Pauli matrices is played by the $\gamma$-matrices.

As shown above, in the chiral case, simple pair exchange is symmetry-forbidden, while twin exchange modifies the mapping. Inspired by the Heisenberg interaction, applying this modification to $\Psi^\xi$ yields the mapping:

\begin{align*}
 \sum_{j} R_j(H_{klmn}) =& \sum_{j} R_j( K(1+\gamma^k\gamma^l)(1+\gamma^m\gamma^n))  \\
=&  4K (1- \epsilon_{klmn}i \gamma^5).
\end{align*}  

This results in a sum of double Heisenberg-like interactions over all $R_j$-allowed vertex permutations (see Appendix \ref{App.Derivation gamma5}). In the chiral tetrahedron the imaginary chirality-dependent contribution arises due to the $\gamma$-matrix algebra  $\gamma^5 \equiv  i\gamma^0\gamma^1\gamma^2\gamma^3 $. Here, $\epsilon_{klmn}$  is +1 for even permutations i.e. a right-handed set of vertices and for odd permutations  it is -1 , i.e. a left-handed set of vertices. 

In mirror-symmetric systems, both even and odd permutations are allowed, ensuring that summing over all possible permutations yields  
$\sum \epsilon_{klmn} i \gamma^5 = 0$,
while the $\gamma^k \gamma^l$ terms remain. Consequently, the system reduces to the Heisenberg description in Eq.~\ref{Eq:h1}. The exchange interaction strength, $K$, can be estimated in organic molecules from the singlet-triplet energy separation to be on the order of 0.1 to 1 eV, much larger than thermal energy at room temperature. 

The additional term, $-iK\gamma^5$, induces non-trivial ordering through twin spin-exchange interactions. Spins align according to the chirality of the system -either pointing inward toward the center of mass or outward—so that the total magnetic moment cancels out (Fig.~\ref{fig:Figure2}c). Any alternative spin configuration on a tetrahedron would yield a net magnetic moment, which is inconsistent with experimental observations and violates $\mathcal{PT}$ symmetry. Under time-reversal, the spin orientation flips (inward $\leftrightarrow$ outward), resulting in a distinct state (bottom to top row in Fig.~\ref{fig:Figure2}c). Parity transforms the tetrahedron into its enantiomer (left to right column), without affecting the spin. Only a combined $\mathcal{PT}$ operation ensures that both enatiomers can be described by the same  Hamiltonian, that is the Hamiltonian that describes chiral systems must obey $\mathcal{PT}$ symmetry. This demonstrates that the spin-exchange interaction in the chiral four-electron system possesses the same $\mathcal{PT}$ symmetry observed in CISS, yet with a non-trivial, chirality-dependent spin texture.

Although Dirac made use of the Lorentz-invariant Clifford algebra omitting the $\gamma^5$-matrix in his original derivation\cite{Dirac1928}, it has been shown that Dirac's theory can be generalized to chiral systems where  $\gamma^5$ also appears \cite{Watson2021}. This is also the only possible generalization that respects Lorentz invariance. We can now write the effective, chiral single particle Dirac equation incorporating the above multi-electron correlations effects in a complex potential as:

\begin{equation}
    (i \hbar \gamma^\mu \partial_\mu - mc - i\frac{K}{c} \gamma^5) \psi = 0
\end{equation}

Assuming all other energy scales are much smaller than the residual mass energy of the electron, so we can restrict ourselves to the non-relativistic limiting case as shown in Appendix \ref{App.NonRelLimit} and resulting in a Schr\"odinger type equation:

\begin{equation}
\left( \frac{\hat{p}^2}{2m}   + \hat{V}  - i \alpha  \hat{\sigma} \cdot \hat{p} \right) \Psi_n = E_n \Psi_n 
\label{Eq. Hamiltonian}
\end{equation}

with $\alpha = \frac{K}{mc}$. 

 The Pauli matrices satisfy the algebra $(\hat{p}\cdot \hat{\sigma})^2 = p^2I$ where $p^2 = p_x^2 + p_y^2 + p_z^2$ and $\hat{p}\cdot \hat{\sigma}$ has the eigenvalues $\pm p$. Adding the $i\hat{p}\cdot \hat{\sigma}$  operator renders the Hamiltonian non-Hermitian. 

The time reversal operation $\mathcal{T} := -i \sigma_y \hat{C}$ where $\hat{C}$ is the complex conjugation operator, fulfills the requirement that the linear and angular momenta are reversed $\mathcal{T}:\hat{p}   \rightarrow -\hat{p} $ and $\mathcal{T}:  \hat{\sigma} \rightarrow -\hat{\sigma} $  while the spatial coordinate is left unchanged $\mathcal{T}:\hat{x}   \rightarrow \hat{x} $ \cite{Dresselhaus2008}. The parity operator $\mathcal{P}$ inverts all spatial coordinates, thus, $\mathcal{P}: \hat{x} \rightarrow -\hat{x}$, for the linear momenta $\mathcal{P}: \hat{p} \rightarrow -\hat{p}$ , but leaves the angular momentum, i.e., the spin orientation, unchanged $\mathcal{P}: \hat{\sigma} \rightarrow \hat{\sigma}$ \cite{Dresselhaus2008}. This is because the linear momentum is a polar vector while the angular momentum is an axial vector\cite{Barron2020}.  

The interaction $i \hat{\sigma}$  is symmetric under $\mathcal{P}: i \hat{\sigma} \rightarrow i \hat{\sigma}$ and $\mathcal{T}: i \hat{\sigma} \rightarrow i \hat{\sigma}$ contrary to the CISS experimental observation, so there is no energetic difference between these four configurations. Only an operator of the form $i \hat{\sigma}\cdot\hat{p}$ breaks $\mathcal{P}$ and $\mathcal{T}$, but remains invariant under the combined $\mathcal{PT}$-operation. 

\begin{align}
    \mathcal{P}:& \quad i \hat{\sigma}\cdot \hat{p} \quad\rightarrow \quad -i \hat{\sigma}\cdot \hat{p} \\
    \mathcal{T}:&\quad i \hat{\sigma}\cdot \hat{p}  \quad\rightarrow \quad  -i \hat{\sigma}\cdot \hat{p} \\
    \mathcal{PT}:& \quad i  \hat{\sigma} \cdot \hat{p} \quad \rightarrow \quad i \hat{\sigma}\cdot \hat{p} 
\end{align}

\section{Analytical 1D model} 
 For the  Schr\"odinger-type Eq.\ref{Eq. Hamiltonian}, there is a solution to an isolated system assuming confinement to one dimension along an arbitrary z-axis. The discretizations in Appendix \ref{App.1DModelDerivation} leads to a digonal banded Hamiltonian matrix, where the up and down spin-components can be block-diagonalized in this case, and one sees that each of the block-diagnoals are Hatano-Nelson models\cite{Hatano1996} for the spin subspaces which are $n\times n$ tri-diagonal Toeplitz matrices.  The solutions to the eigenvalue problems gives double degenerated, spin-independent energies 

\begin{equation}
    E_{j} = V_0 + \frac{\hbar^2}{m\Delta z^2}+\hbar \frac{ \sqrt{\hbar^2 - \alpha^2 m^2 \Delta z^2}}{m\Delta z^2} \cos{ \frac{ j \pi}{n+1} } .
\end{equation}

\begin{figure*}
\includegraphics[width = 0.90\textwidth]{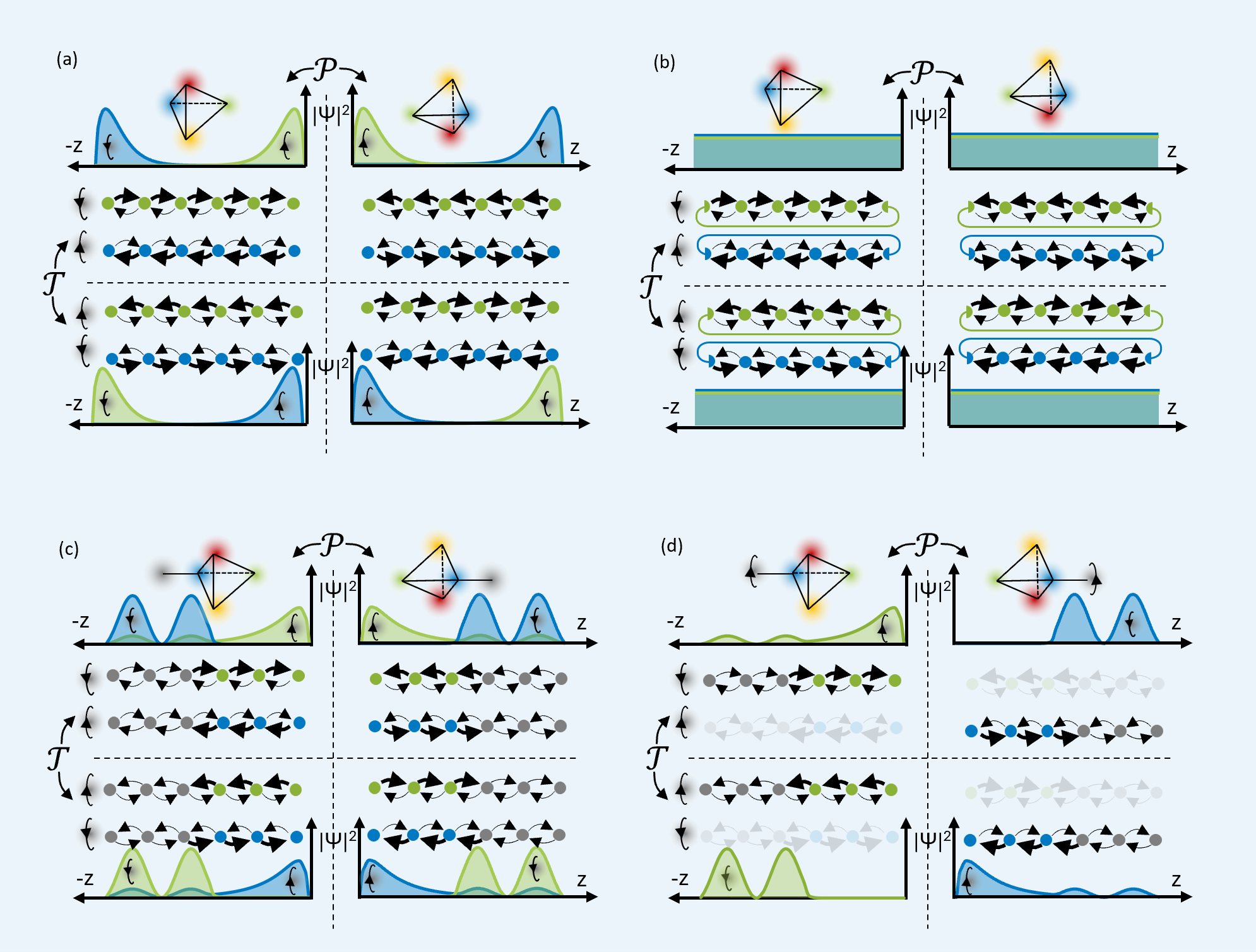}
\caption{ Illustration of the model outcomes for a chiral material with open boundary conditions (a), for chiral material with periodic boundary conditions (b),  for a chiral material linked to a non-chiral material (c) and for a chiral material linked to a ferro-magnetic material (d). In case of periodic boundary conditions (b) the states are perfectly delocalized. For cases (a-c) two chiral states are degenerate in energy - but for (d, the extended state is energetically more favorable than the isolated state on the ferromagnet. In all the cases, applying $\mathcal{PT}$ yields a degenerate energy state (on the opposite enantiomer). }
    \label{fig:Figure3}
\end{figure*}

Figure~\ref{fig:Figure3}a shows the spatial distribution of the probability density of the right eigenvectors that are localized on either interface depending on spin. The right eigenvector for the spin-up state is
\begin{equation}
    \Psi_{j,{\uparrow}} (k) = A  e^{-k \ln{\sqrt{ \frac{\hbar^2 - \alpha \hbar m \Delta z}{\hbar^2 + \alpha \hbar m \Delta z}  } }} \sin{ \frac{ j k \pi}{n+1} } 
\end{equation}

 whereas the right eigenvector for the spin down-state is
 \begin{equation}
    \Psi_{j,{\downarrow}} (k) = A  e^{-k \ln{\sqrt{ \frac{\hbar^2 + \alpha \hbar m \Delta z}{\hbar^2 - \alpha \hbar  m \Delta z}  } }} \sin{ \frac{ j k \pi}{n+1} }
\end{equation}

with $ k=1,2,...,n$. The left eigenvectors derive from the right eigenvectors as $\langle \psi| = \mathcal{P}|\psi \rangle$.

Energies are real as long as $\hbar > |\alpha \Delta z m|$ which is a common feature for $\mathcal{PT}$-systems\cite{Bender2007}. The use of this critical point gives an upper limit of the interaction strength of $K \approx 2\cdot10^2$ eV. The solutions are self-consistent with the  chiral spin-texture discussed in Fig. \ref{fig:Figure2},  i.e., the resulting states are $\mathcal{PT}$ symmetric where the parity operation acts on  $ k  \rightarrow  k-n-1$ and $ \Delta z  \rightarrow  -\Delta z$ , and  results in wavefunctions that display the  opposite spin localization (Fig. \ref{fig:Figure3}a left panel to right panel) i.e. the time-reversed spin-texture and the orginial wavefunction is recovered by applying the time-reversal operation (Fig. \ref{fig:Figure3}a top to bottom panel). The $\mathcal{P}$ operator for this particular example is a  $2n\times2n$-matrix where $n$ is the number of sites and it is

\begin{equation}
  \mathcal{P} = \begin{bmatrix}
0 & 0 & \cdots &0 &  I\\
0 & 0 & \cdots & I &  0 \\
\vdots & \vdots & \reflectbox{$\ddots$} & \vdots &  \vdots\\
0 & I & \cdots &0 &  0\\
I & 0 & \cdots & 0 &  0 \\
\end{bmatrix}
\end{equation}

where $I$ is the $2\times2$ identity matrix with the property $\mathcal{P}=\mathcal{P}^{-1}$ with eigenvalues $\pm1$. One can see that $\mathcal{P}$  defined above reverses the order of the $n$ sites, but leaves the spin unchanged. The time-reversal operator is defined as $\mathcal{T}_n := -i \sigma_2 \hat{C} \otimes I_n$,  and thus it reverses all the spins but the ordering of the sites remains invariant. 

The non-Hermitian skin effect refers to the accumulation of waves or particles at system boundaries due to asymmetric hopping~\cite{Kunst2018}. This directional imbalance disrupts the usual bulk-boundary correspondence of Hermitian systems, where open and periodic boundary conditions yield similar bulk properties. In non-Hermitian systems, however, Bloch band theory fails~\cite{Kunst2018}.  This localization contrasts sharply with Hermitian systems, where, under open boundary conditions, a sufficiently large system smoothly transitions to the band structure with periodic boundary conditions. In non-Hermitian systems, this bulk-boundary correspondence and Bloch band theory break down~\cite{Kunst2018}. In above non-Hermitian model, connecting the interfaces restores periodic boundary conditions. The corresponding eigenvalues form a circle in the complex energy plane, with each spin species exhibiting two counter-propagating modes with real eigenvalues, leading to fully delocalized states (Fig.~\ref{fig:Figure3}b). While spin-up and spin-down states could, in principle, couple—weakening the skin effect, they would still localize at opposite boundaries under open conditions. 

Spin and charge localization also arise when a structurally chiral domain is coupled to Hermitian leads, such as metallic, diamagnetic, or ferromagnetic contacts common in CISS experiments. In a simplified model, diamagnetic leads exhibit symmetric hopping and spin-independent on-site energies, while ferromagnetic leads retain symmetric hopping but introduce spin-dependent energy splitting.

Figure~\ref{fig:Figure3}c shows the probability density for a non-magnetic lead with opposite enantiomers. Preferential hopping within the chiral domain induces spin accumulation (\ref{fig:Figure3}c, green wavefunction, top panel) or expulsion (\ref{fig:Figure3}c, blue wavefunction, top panel), generating local spin imbalance and magnetization extending into the non-chiral material. However, the net magnetization remains zero since no net spin imbalance exists across the entire system. In practice, spin polarization in non-chiral materials persists only within the spin-decay length within the non-chiral material, localizing magnetization at the interface.

Figure~\ref{fig:Figure3}d depicts probability densities for a magnetic lead with opposite enantiomers (left to right) and opposite lead magnetization (top to bottom). Here, energetic preference for one spin species in the ferromagnetic lead results in majority spin occupation. As in the metallic case, the wavefunction is either attracted to or repelled from the chiral domain. However, the single-spin occupation leads to two distinct energy states: a lower-energy delocalized state (\ref{fig:Figure3}d, green wavefunction, top panel; blue, bottom panel) and a confined state excluding spins from the chiral domain (\ref{fig:Figure3}d, blue wavefunction, top panel; green, bottom panel). While minority carriers exist in reality, the qualitative result remains unchanged, producing two pairs of energetically equivalent states on the diagonals in Fig. \ref{fig:Figure3}d.

A natural extension involves connecting the chiral domain to semi-infinite reservoirs to study transport properties. As seen previously, charge transport through the chiral domain is spin-polarized due to spin-momentum locking, implying that pure spin transport also induces spin-polarized charge transport. This is also backed with an scattering argument in Appendix \ref{Scattering}. However, analyzing transport in this non-Hermitian system is beyond the scope of this work, as standard methods like the Landauer-Büttiker formalism and non-equilibrium Green’s functions methods rely on Hermiticity~\cite{Ashida2020}.

\section{Discussion}

With the established model, the stage is now set to interpret a broad range of CISS-related experimental phenomena. All observed CISS effects fall within the scope of this model, beginning with quasi-static responses such as induced magnetization and electric polarization, and extending to charge and spin transport phenomena.

\subsection{Quasi-static CISS observations}
The model does not rely on specific geometries, as the mere absence of mirror symmetry and the presence of electronic spin are sufficient. Consequently, CISS effects should be observable along crystallographic axes lacking screw symmetry, provided the material is chiral, independent of specific potential landscapes such as helical potentials. However, different scales of extension dimension can affect the strength of spin localization and therefore the strength of CISS. %CitationAzim'sPaper?

Using Hall-voltages on a  two-dimensional electron gas in GaAs or gold with absorbed chiral peptides show the ability of the chiral structure to transfer spins into a diamagnetic material\cite{Smolinsky2019}. This aligns well with our model prediction for the diamagnetic-chiral material junction in Figure \ref{fig:Figure3}b. 

Similarly chiral/ferromagnetic junctions can induce magnetization switching in the ferromagnet~\cite{BenDor2017} as illustrated in Figure \ref{fig:Figure3}c. The resulting spin polarization is strong enough to stabilize a super-paramagnetic layer and this is due to the spin-exchange interaction and not spin-orbit coupling. The model predicts the observations such as shifts in the Curie temperature\cite{Koplovitz2019}, alters the magnetization\cite{Meirzada2021,BenDor2017} or modifications to the coercive field\cite{Koplovitz2019}, and that two ferromagnetic layers on opposite sides of a chiral layer should polarize in opposing directions. Opposite spin polarization at interfaces has been observed by Nakajima et al.~\cite{Nakajima2023}. These predictions are based on the coupling between a chiral material to a ferromagnetic material in Figure  \ref{fig:Figure3}c.  Flipping the enantiomer toggles the magnetic polarization due to the $\mathcal{PT}$ symmetry. This contrasts with the Edelstein effect, where opposite interfaces exhibit identical spin polarization. 

Since charge and spin are momentum-locked, not only a spin polarization accompanies chiral materials, but also a static charge polarization. In case of a ferromagnet/chiral junction, the different spin-charge localizations at interfaces alters screening behavior, leading to differences in charge distribution and bandbending between states of opposite magnetization or opposite enantiomers. This difference can be measured as an change in the surface potential\cite{Abendroth2019,Ghosh2020}, modulation of the electric field at the Shottky junction resulting in difference is tunneling barrier height depending on magnetization of chirality or as a chirality-induced quantum capacitance, as observed in Kelvin probe force measurements~\cite{Theiler2023}.

Spin localization modulates the interaction between the ferromagnetic and opposite enantiomers. This also can explain chiral recognition at magnetized surfaces where a preferred enantiomer is absorbed depending on the chirality type and magnetization direction\cite{Koyel2018} . Spin-exchange force observed in AFM experiments is evidence for a new enantiospecific interaction force in chiral bio-molecules which presumes a spin-polarization a source of the effect\cite{Kapon2021}. In our model presented here this spin polarization is provided by  spin-localization at the boundary of the chiral system. Generally, exchange interactions in metals are weaker than in insulators since because of the increased screening and delocalization of the wavefunction.  

The proposed model can be easily extended to include additional effects such as temperature-related interactions. These effects can soften the strong spin polarization, but do not change the basic symmetric argumentation. The result of these considerations is that, for example, spin-polarization in a non-chiral material will only be transferred up to the spin-diffusion length within that corresponding non-chiral material. This has been observed in experiments in Kelvin Probe force microcopy experiments where the shift of local contact potential decreases as the thickness of buffer layer between the chiral material and the ferromagnet increases \cite{Ghosh2020}.

\subsection{Transport related CISS observations}

For the observation of CISS enhancement in (electro-) catalysis, particularly for triplet-ground-state molecules like molecular oxygen, a spin polarized electron current provides the necessary angular momentum needed for the reaction so that directly the ground state molecules can be formed\cite{Namaan2024}. Using the model of an electrode in contact with chiral material (Fig.\ref{fig:Figure3}b), the model shows that spin polarization through the localized boundary effect can reach very high levels. A spin-polarized current can also be created in absence of magnetic interface as shown in the chiral layer spin-LED \cite{Hautzinger2024}. Since charge transfer in electrocatalysis requires sufficient orbital overlap, only spin-polarized species contribute, while the opposite spin species remain buried in the bulk or are chemically inactive at the interface.  This is in contrast to ferromagnetic materials, there majority and minority carriers are present at the interface.  Since spin polarization is present also in absence of a current, this explains why  Liang et al.~\cite{Liang2022a} have observed boosted activity even when there is no direct current flow through the chiral material.

For magneto-resistive observations, we need to add transport to a situation like in Fig.\ref{fig:Figure3}b. Magneto-resistive experiments reveal different conductivities for opposite magnetization at zero bias\cite{Al-Bustami2022}, indicating an underlying mechanism active in thermodynamic equilibrium and in the absence of global charge transport. The proposed mechanism fulfills these conditions. This effect persists in both closed and open systems at equilibrium. Bardarson's theorem~\cite{Bardarson2008} remains valid, as outlined here in this work,  with spin, time-reversal symmetry is broken alongside spatial symmetry. The conservation of combined $\mathcal{PT}$-symmetry ensures real eigenvalues, leading to stable systems. A reformulation of Bardarson's theorem suggests that in a $\mathcal{PT}$ symmetric system ($[\mathcal{PT},H] =0$), the eigen-energies still come in pairs. However since $H(\mathcal{PT}\Psi) = \mathcal{PT}(H\Psi) = \mathcal{PT}(E\Psi) = E^*(\mathcal{PT} \Psi)$, $\mathcal{PT}\Psi$ is the degenerate partner of $\Psi$ on the other chiral sector. Consequently, spin-polarized transport is possible in two-terminal devices without dissipation in chiral structures. The analogous mechanism of CISS quantum capacitance can also be used in transport measurements to explain the alteration of the Schottky barrier at a chiral/ferromagnetic interface under enantiomer inversion or magnetization reversal~\cite{Tirion2024} which can be applied to describe the high magneto-resistance of chirality-induced tunneling magneto-resistance devices\cite{Al-Bustami2022}.

The transport through a ferro-magnetic- chiral material junction appears as well in photoemission experiments\cite{Abendroth2019}. Energetic photons provided the energy to induce electron emission through the junction. Interestingly, the experiment could also detect a change in work function\cite{Abendroth2019}. In the absence of a ferromagnetic component, circular polarized light can provide photo-excited spin-polarized electrons, introducing a spin-imbalance similar to of a ferromagnet interface, which leads to observation of the CISS photoelectron emission\cite{Mollers2022}. Experiments on donor-chiral-bridge-acceptor molecules show that there is no need for metal/chiral material interface needed to observe CISS\cite{Eckvahl2023}, the static magnetic field is enough to break the symmetry to be detected in the electron-paramagnetic resonance measurements. This aligns well with the description that only an open-boundary condition of the chiral material is needed to create energetically degenerate spin-imbalance at the two ends of the chiral molecule.

The inverse effect of CISS\cite{Inui2020,Sun2024}, where a pure spin current induces a charge current can also described with the proposed model. A pure spin current that accumulates inside a chiral material leads to an imbalance of the equilibrium spin balance, to compensate a spin-momentum locked charge current occurs. In further spin-pumping experiments\cite{Moharana2025,Sun2024_1}, the alignment between spin and linear momentum of the electron in chiral material was detected. This is consistent with the postulated $i \alpha \hat{p}\cdot \hat{\sigma}$ CISS term to account for the chiral symmetry discussed above.

\subsection{New directions}

The distinction between bulk and interface effects in CISS becomes obsolete within this framework. Both are required to explain the full range of phenomena. Bulk symmetry breaking gives rise to spin-momentum-locked states, while interfaces localize these states via the non-Hermitian skin effect. Unlike Hermitian systems, where periodic and open boundary conditions yield similar bulk spectra apart from edge states, non-Hermitian systems exhibit a complete restructuring of their spectral and topological properties~\cite{Kunst2018}. In the proposed model, periodic boundary conditions produce a loop of complex eigenvalues and delocalized states. Open boundaries collapse the spectrum to a real segment, localizing all states at the interfaces. Crucially, most conventional simulation tools—based on periodicity and Hermiticity—are blind to these effects. Non-Hermitian dynamics and many-body correlations are essential but computationally demanding, requiring new theoretical approaches tailored to structurally chiral systems.

Artificial chiral molecules engineered with ultracold atoms offer a tunable platform\cite{Gross2017} to test the exchange-driven mechanism underlying CISS. By arranging spinful atoms in tetrahedral geometries, it is possible to simulate twin-pair exchange interactions and $\mathcal{PT}$-symmetric non-Hermitian dynamics. A particularly revealing configuration is a ring of chiral units with a tunable weak link. This geometry continuously interpolates between a rod with open boundaries and a ring with periodic boundary conditions. As the link is gradually closed, the system transitions from strong boundary localization—characteristic of the non-Hermitian skin effect—to delocalized bulk-like behavior. Time-resolved measurements of spin and density correlations across this transition can directly probe the interplay between non-Hermiticity, symmetry breaking, and structural chirality\cite{Gross2017}. This approach offers a powerful route to experimentally test the theoretical model and its predicted breakdown of bulk-boundary correspondence.

\section{Conclusion}

CISS has challenged prevailing paradigms in spin transport, quantum chemistry, and biophysics. This work introduces a unifying theoretical framework that identifies structural chirality as the origin of spin-momentum locking via twin-pair exchange interactions governed by non-Hermitian quantum statistics. Crucially, chirality is not a local or single-particle property—it emerges as a non-local phase constraint in multi-electron systems, inherently tied to the spin-statistics theorem and the absence of mirror symmetry. These constraints break both parity and time-reversal symmetry, yet enforce a combined $\mathcal{PT}$ symmetry that stabilizes the system with real eigenvalues. The result is a self-consistent mechanism for spin selectivity, rooted in symmetry and quantum correlations, that does not require strong spin-orbit coupling or magnetic fields.

This approach reveals structural chirality as a generator of non-Hermitian topology, leading to interface-localized spin accumulation and energetically distinct spin-filtered states. These effects depend jointly on bulk symmetry breaking and boundary-induced localization—features inaccessible to conventional computational models that rely on periodic boundary conditions or Hermitian assumptions. The findings thus call for a reevaluation of simulation strategies, especially for complex, correlated systems where structural chirality is present.

Beyond physics, the implications are far-reaching. In chemistry, the model offers a natural explanation for spin-dependent reactivity in asymmetric catalysis. In biology, it lends quantitative support to theories linking homochirality, metabolism, and enzymatic selectivity to spin-correlated electron dynamics. The emergence of spin selectivity from symmetry, rather than from energetic bias or external fields, may point to a quantum-statistical origin of life's molecular asymmetry.

In quantum technology, the framework paves the way for initializing, manipulating, and reading out spin states at room temperature—potentially enabling robust quantum sensing and thermally driven spintronic devices. The inherent non-reciprocity introduced by chirality opens new avenues for designing quantum ratchets, directional transport systems, and information engines powered by structural asymmetry rather than energetic gradients.

Together, these insights suggest a paradigm shift: chirality must be recognized as a fundamental quantum degree of freedom that reshapes how spin, structure, and symmetry interact. From understanding the origins of life to designing the next generation of quantum materials and devices, chirality offers an axis along which physical laws and technological innovations may evolve.

\begin{acknowledgments}
The authors gratefully acknowledge Volker Blum and Peter C. Sercel for their insightful comments and fruitful discussions.
\textbf{Funding: }This project was supported as part of the Center for Hybrid Organic Inorganic Semiconductors for Energy (CHOISE) an Energy Frontier Research Center funded by the Office of Basic Energy Sciences, Office of Science within the U.S. Department of Energy. This work was authored in part by NREL under Contract No. DE-AC36-08GO28308 to DOE. The views expressed in the article do not necessarily represent the views of the DOE or the U.S. Government.\\
\textbf{Competing interests:} The authors declare no competing interests. \\
\textbf{Data availability:} The main data that support the findings of this study are available in this article. Additional data are available from the corresponding authors upon reasonable request. \\

\end{acknowledgments}

\appendix

\section{Derivation of the $\gamma^5$ Potential}
\label{App.Derivation gamma5}
For the right handed, chiral tetrahedron, the four generators for the permutation group yields for $R_1$:

\begin{align}
H_{0123} =& K(1 + \gamma^0 \gamma^1 + \gamma^2 \gamma^3 + \gamma^0 \gamma^1 \gamma^2 \gamma^3) \\ =& K( 1 + \gamma^0 \gamma^1 + \gamma^2 \gamma^3 - i \gamma^5) 
\end{align}

for $R_2$;
\begin{align}
H_{3210} =& K(1 + \gamma^3\gamma^2 + \gamma^1 \gamma^0 + \gamma^3 \gamma^2 \gamma^1 \gamma^0) \\ =& K( 1 + \gamma^3 \gamma^2 + \gamma^1 \gamma^0 - i \gamma^5) 
\end{align}

for $R_3$;
\begin{align}
H_{1032} =& K(1 + \gamma^1\gamma^0 + \gamma^3 \gamma^2 + \gamma^1 \gamma^0 \gamma^3 \gamma^2) \\ =& K( 1 + \gamma^1\gamma^0 + \gamma^3 \gamma^2 - i \gamma^5) 
\end{align}

for $R_4$;
\begin{align}
H_{2301} =& K(1 + \gamma^2\gamma^3 + \gamma^0 \gamma^1 + \gamma^2 \gamma^3 \gamma^0 \gamma^1) \\ =& K( 1 + \gamma^2 \gamma^3 + \gamma^0 \gamma^1 - i \gamma^5) 
\end{align}

Using the property that  $\gamma^0\gamma^1 = -\gamma^1\gamma^0 $  and  $\gamma^2\gamma^3 = -\gamma^3\gamma^2 $, one can find the sum over all allowed permutations of right-handed, chiral tetrahedron is

\begin{align}
 \sum_{j} R_j(H_{klmn}) =& H_{0123} + H_{3210} + H_{1032} + H_{2301} \\ =& 4K (1 - i\gamma^5)
\end{align}

Similarly, for the chiral, left-handed, chiral tetrahedron one finds
\begin{align}
 \sum_{j} R_j(H_{klmn}) =& H_{1023} + H_{3201} + H_{0132} + H_{2310} \\ =& 4K (1 + i\gamma^5)
\end{align}

\section{Non relativistic limit of Dirac equation}
\label{App.NonRelLimit}

The chiral Dirac equation can be written in its time-independent $H \Psi_n = E_n \Psi_n$ for an additional electrostatic potential $\hat{V}$ in a 2-spinor form with $\Psi_n = (\Psi^L_n,  \Psi^S_n)^T $  with large and small components and the Hamiltonian is

\begin{equation}
 H = \begin{pmatrix} 
 \hat{V}  & c \hat{\sigma} \cdot \hat{p} \\
 c \hat{\sigma} \cdot \hat{p} - 2i K & \hat{V} - 2mc^2 \\
 \end{pmatrix}   
\end{equation}

The term \( \hat{\sigma} \cdot \hat{p} \) represents the inner product between the Pauli vector \( \hat{\sigma} = (\sigma_x, \sigma_y, \sigma_z) \) and the momentum operator \( \hat{p} = (\hat{p}_x, \hat{p}_y, \hat{p}_z) \), expressed as: $\hat{\sigma} \cdot \hat{p} = \sigma_x \hat{p}_x + \sigma_y \hat{p}_y + \sigma_z \hat{p}_z.$ The small component can be expressed as 

\begin{equation}
 \Psi^S_n = \frac{c \hat{\sigma} \cdot \hat{p} - 2i K   }{E_n -\hat{V} + 2mc^2} \Psi^L_n
\end{equation}

Substituting into top equations yields an equation for the equation of the large eq:
\begin{equation}
\left(  \hat{V}  + c \hat{\sigma} \cdot \hat{p} \frac{\hat{Q}}{2mc^2} c \hat{\sigma} \cdot \hat{p}  - i \hat{\sigma} \cdot \hat{p} \frac{K}{c} \hat{Q} \right) \Psi^L_n = E_n \Psi^L_n
\end{equation}

with 
\begin{equation}
\hat{Q} = \left( 1 + \frac{E_n - V}{2mc^2 } \right)^{-1} \approx  1 - \frac{E_n - V}{2mc^2}
\end{equation}

using the approximation in non-relativistic limits yields

\begin{align*}
&\left( \frac{\hat{p}^2}{2m }  + \hat{V}  - \frac{c \hat{\sigma} \cdot \hat{p} (E_n - \hat{V}) c \hat{\sigma} \cdot \hat{p}}{4m^2c^4 }  \right. \dots \\ 
&\left. - i \hat{\sigma} \cdot \hat{p} \frac{K}{mc} \left(1 - \frac{E_n - V}{2mc^2 }))\right)\right) \Psi^L_n = E_n \Psi^L_n \\
\end{align*}

where the first two terms correspond to the usual Schr\"odinger equation, the third term incorporates  the relativistic corrections including spin-orbit coupling and the last term is the chirality-induced spin selectivity term needed for the twin-pair exchange discussed above. The relevant energy scale for electrons in chiral molecules is much smaller than the rest mass of the electron of 510 keV,  and here we  neglect all relativistic corrections and replace $\alpha = \frac{K}{mc}$ resulting in 

\begin{equation}
\left( \frac{\hat{p}^2}{2m}   + \hat{V}  - i\alpha \hat{\sigma} \cdot \hat{p} \right) \Psi^L_n = E_n \Psi^L_n 
\end{equation}

\section{Analytical 1D model derivations} 
\label{App.1DModelDerivation}
From Eq. \ref{Eq. Hamiltonian}, we find the general Hamiltonian that breaks both $\mathcal{P}$  and $\mathcal{T}$ but not $\mathcal{PT}$ symmetries is

\begin{equation}
    H = \frac{\hat{p}^2}{2m} + \hat{V} - i \alpha \hat{p}\cdot \hat{\sigma} 
\end{equation}

In one dimension, the problem can be simplified. Assuming confinement in one dimension along an arbitrary z axis, we can write Eq. \ref{Eq. Hamiltonian}

\begin{figure}
\includegraphics[width = 0.45\textwidth]{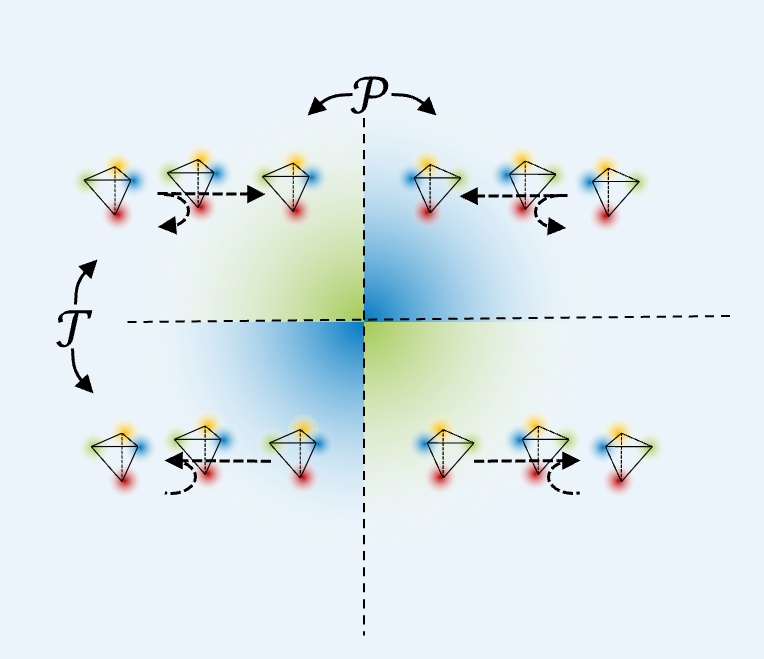}
\caption{ Illustration of a scattering process of chiral particles under time-reversal and parity symmetry operations.     }
    \label{fig:Figure4}
\end{figure}

\begin{equation}
   \left( -\sigma_0 \frac{\hbar^2}{2m} \frac{\partial^2}{\partial z^2} + \sigma_z \alpha \hbar \frac{\partial}{\partial z  } +V_0 \right) \Psi(x) = E \Psi(x)
\end{equation} 

This continuous particle-in-box model can be transformed into a discretized tight-binding form,
\begin{equation}
     \frac{\partial}{\partial z} \Psi \approx \frac{\Psi_{j+1}-\Psi_{j-1}}{2 \Delta z}
\end{equation}
and, 
\begin{equation}
     \frac{\partial^2}{\partial z^2} \Psi \approx \frac{\Psi_{j+1}-2\Psi_{j}+\Psi_{j-1}}{\Delta z^2}
\end{equation}

where one can write for the $j$-th spin up $\Psi_{\uparrow,j}$  and spin down $\Psi_{\downarrow,j}$

\begin{equation}
\begin{bmatrix}
a & 0 & b & 0 & 0 & 0 & \cdots &  0 \\
0 & a & 0 & c & 0 & 0 & \cdots &  0 \\
c & 0 & a & 0 & b & 0 & \cdots &  0 \\
0 & b & 0 & a & 0 & c & \cdots &  0 \\
\vdots &  & \ddots &  & \ddots & &\ddots & \vdots \\
0 & \dots & 0 & b& 0 & a & 0 &  c \\
0 & \dots & 0 & 0 & c& 0 & a &  0 \\
0 & \dots & 0 & 0 & 0 & b & 0 &  a \\

\end{bmatrix} \begin{pmatrix}
\Psi_{\uparrow1} \\
\Psi_{\downarrow1} \\
\Psi_{\uparrow2} \\
\Psi_{\downarrow2} \\
\vdots \\
\vdots \\
\Psi_{\uparrow n} \\
\Psi_{\downarrow n} \end{pmatrix} \\ = E \begin{pmatrix}
\Psi_{\uparrow1} \\
\Psi_{\downarrow1} \\
\Psi_{\uparrow2} \\
\Psi_{\downarrow2} \\
\vdots \\
\vdots \\
\Psi_{\uparrow n} \\
\Psi_{\downarrow n} \\
\end{pmatrix}
\end{equation}

with 

\begin{align*}
    a &= V_0 + \frac{\hbar^2}{m \Delta z^2 }   \\ b &= \frac{\hbar^2 + \alpha \hbar  m \Delta z }{2m\Delta z^2}  \\ c &= \frac{\hbar^2 - \alpha \hbar  m \Delta z }{2m\Delta z^2} .
\end{align*}

This is a diagonal banded Hamiltonian matrix, where the up and down spin-components can be block-diagonalized, and one sees that each of the block-diagonals are Hatano-Nelson models\cite{Hatano1996} for the spin subspaces which are $n\times n$ tri-diagonal Toeplitz matrices.

\section{Chiral Scattering}
\label{Scattering}
Chiral system can also be analyzed as a scattering problem in dynamic equilibrium, subject to non-trivial time and spatial symmetry constraints. Lifshitz \cite{Lifshitz1981} noted that chiral scatterers require a special form of detailed balance. Figure \ref{fig:Figure4} illustrates the collision process of chiral identical particles, $C_v + C_w \rightarrow C_{v'} + C_{w'}$, with initial velocities $v$ and $w$ and final velocities $v'$ and $w'$. Under time reversal, the reversed collision follows as $C_{-v'} + C_{-w'} \rightarrow C_{-v} + C_{-w}$. However, the sequence of events is not exactly reversed: in the forward process, the left particle in Figure \ref{fig:Figure2} sees a sequence of blue, then green during the collision while in the time-reversed process, the sequence remains blue, then green.  
To satisfy detailed balance, the reaction must instead follow $C_{v'} + C_{w'} \rightarrow C_v + C_w$, requiring an additional $\mathcal{P}$ transformation to restore the correct sequence of first green, then blue scattering. As a result, in a chiral system, the scattering probabilities for forward and reverse processes may differ and are invariant under $\mathcal{PT}$-symmetry\cite{Bender2007}. To achieve dynamic equilibrium, where scattering rates are equal, the system must adjust by developing different concentrations of particles.  

A similar principle is well-known in chemistry: a reaction in solution can reach dynamic equilibrium, where no macroscopic flow is observed because the inflow and outflow of reaction species are balanced. According to the law of mass action, if the reaction rates are initially unequal, the concentrations adjust until the flows equalize.


\begin{thebibliography}{44}%
\makeatletter
\providecommand \@ifxundefined [1]{%
 \@ifx{#1\undefined}
}%
\providecommand \@ifnum [1]{%
 \ifnum #1\expandafter \@firstoftwo
 \else \expandafter \@secondoftwo
 \fi
}%
\providecommand \@ifx [1]{%
 \ifx #1\expandafter \@firstoftwo
 \else \expandafter \@secondoftwo
 \fi
}%
\providecommand \natexlab [1]{#1}%
\providecommand \enquote  [1]{``#1''}%
\providecommand \bibnamefont  [1]{#1}%
\providecommand \bibfnamefont [1]{#1}%
\providecommand \citenamefont [1]{#1}%
\providecommand \href@noop [0]{\@secondoftwo}%
\providecommand \href [0]{\begingroup \@sanitize@url \@href}%
\providecommand \@href[1]{\@@startlink{#1}\@@href}%
\providecommand \@@href[1]{\endgroup#1\@@endlink}%
\providecommand \@sanitize@url [0]{\catcode `\\12\catcode `\$12\catcode `\&12\catcode `\#12\catcode `\^12\catcode `\_12\catcode `\%12\relax}%
\providecommand \@@startlink[1]{}%
\providecommand \@@endlink[0]{}%
\providecommand \url  [0]{\begingroup\@sanitize@url \@url }%
\providecommand \@url [1]{\endgroup\@href {#1}{\urlprefix }}%
\providecommand \urlprefix  [0]{URL }%
\providecommand \Eprint [0]{\href }%
\providecommand \doibase [0]{https://doi.org/}%
\providecommand \selectlanguage [0]{\@gobble}%
\providecommand \bibinfo  [0]{\@secondoftwo}%
\providecommand \bibfield  [0]{\@secondoftwo}%
\providecommand \translation [1]{[#1]}%
\providecommand \BibitemOpen [0]{}%
\providecommand \bibitemStop [0]{}%
\providecommand \bibitemNoStop [0]{.\EOS\space}%
\providecommand \EOS [0]{\spacefactor3000\relax}%
\providecommand \BibitemShut  [1]{\csname bibitem#1\endcsname}%
\let\auto@bib@innerbib\@empty
%</preamble>
\bibitem [{\citenamefont {Bloom}\ \emph {et~al.}(2024)\citenamefont {Bloom}, \citenamefont {Paltiel}, \citenamefont {Naaman},\ and\ \citenamefont {Waldeck}}]{Bloom2024}%
  \BibitemOpen
  \bibfield  {author} {\bibinfo {author} {\bibfnamefont {B.~P.}\ \bibnamefont {Bloom}}, \bibinfo {author} {\bibfnamefont {Y.}~\bibnamefont {Paltiel}}, \bibinfo {author} {\bibfnamefont {R.}~\bibnamefont {Naaman}},\ and\ \bibinfo {author} {\bibfnamefont {D.~H.}\ \bibnamefont {Waldeck}},\ }\href@noop {} {\bibfield  {journal} {\bibinfo  {journal} {Chemical Reviews}\ }\textbf {\bibinfo {volume} {124}},\ \bibinfo {pages} {1950} (\bibinfo {year} {2024})}\BibitemShut {NoStop}%
\bibitem [{\citenamefont {Aiello}\ \emph {et~al.}(2022)\citenamefont {Aiello}, \citenamefont {Abendroth}, \citenamefont {Abbas}, \citenamefont {Afanasev}, \citenamefont {Agarwal}, \citenamefont {Banerjee}, \citenamefont {Beratan}, \citenamefont {Belling}, \citenamefont {Berche}, \citenamefont {Botana}, \citenamefont {Caram}, \citenamefont {Celardo}, \citenamefont {Cuniberti}, \citenamefont {Garcia-Etxarri}, \citenamefont {Dianat}, \citenamefont {Diez-Perez}, \citenamefont {Guo}, \citenamefont {Gutierrez}, \citenamefont {Herrmann}, \citenamefont {Hihath}, \citenamefont {Kale}, \citenamefont {Kurian}, \citenamefont {Lai}, \citenamefont {Liu}, \citenamefont {Lopez}, \citenamefont {Medina}, \citenamefont {Mujica}, \citenamefont {Naaman}, \citenamefont {Noormandipour}, \citenamefont {Palma}, \citenamefont {Paltiel}, \citenamefont {Petuskey}, \citenamefont {Ribeiro-Silva}, \citenamefont {Saenz}, \citenamefont {Santos}, \citenamefont {Solyanik-Gorgone}, \citenamefont {Sorger}, \citenamefont {Stemer}, \citenamefont
  {Ugalde}, \citenamefont {Valdes-Curiel}, \citenamefont {Varela}, \citenamefont {Waldeck}, \citenamefont {Wasielewski}, \citenamefont {Weiss}, \citenamefont {Zacharias},\ and\ \citenamefont {Wang}}]{Aiello2022}%
  \BibitemOpen
  \bibfield  {author} {\bibinfo {author} {\bibfnamefont {C.~D.}\ \bibnamefont {Aiello}}, \bibinfo {author} {\bibfnamefont {J.~M.}\ \bibnamefont {Abendroth}}, \bibinfo {author} {\bibfnamefont {M.}~\bibnamefont {Abbas}}, \bibinfo {author} {\bibfnamefont {A.}~\bibnamefont {Afanasev}}, \bibinfo {author} {\bibfnamefont {S.}~\bibnamefont {Agarwal}}, \bibinfo {author} {\bibfnamefont {A.~S.}\ \bibnamefont {Banerjee}}, \bibinfo {author} {\bibfnamefont {D.~N.}\ \bibnamefont {Beratan}}, \bibinfo {author} {\bibfnamefont {J.~N.}\ \bibnamefont {Belling}}, \bibinfo {author} {\bibfnamefont {B.}~\bibnamefont {Berche}}, \bibinfo {author} {\bibfnamefont {A.}~\bibnamefont {Botana}}, \bibinfo {author} {\bibfnamefont {J.~R.}\ \bibnamefont {Caram}}, \bibinfo {author} {\bibfnamefont {G.~L.}\ \bibnamefont {Celardo}}, \bibinfo {author} {\bibfnamefont {G.}~\bibnamefont {Cuniberti}}, \bibinfo {author} {\bibfnamefont {A.}~\bibnamefont {Garcia-Etxarri}}, \bibinfo {author} {\bibfnamefont {A.}~\bibnamefont {Dianat}}, \bibinfo {author}
  {\bibfnamefont {I.}~\bibnamefont {Diez-Perez}}, \bibinfo {author} {\bibfnamefont {Y.}~\bibnamefont {Guo}}, \bibinfo {author} {\bibfnamefont {R.}~\bibnamefont {Gutierrez}}, \bibinfo {author} {\bibfnamefont {C.}~\bibnamefont {Herrmann}}, \bibinfo {author} {\bibfnamefont {J.}~\bibnamefont {Hihath}}, \bibinfo {author} {\bibfnamefont {S.}~\bibnamefont {Kale}}, \bibinfo {author} {\bibfnamefont {P.}~\bibnamefont {Kurian}}, \bibinfo {author} {\bibfnamefont {Y.-C.}\ \bibnamefont {Lai}}, \bibinfo {author} {\bibfnamefont {T.}~\bibnamefont {Liu}}, \bibinfo {author} {\bibfnamefont {A.}~\bibnamefont {Lopez}}, \bibinfo {author} {\bibfnamefont {E.}~\bibnamefont {Medina}}, \bibinfo {author} {\bibfnamefont {V.}~\bibnamefont {Mujica}}, \bibinfo {author} {\bibfnamefont {R.}~\bibnamefont {Naaman}}, \bibinfo {author} {\bibfnamefont {M.}~\bibnamefont {Noormandipour}}, \bibinfo {author} {\bibfnamefont {J.~L.}\ \bibnamefont {Palma}}, \bibinfo {author} {\bibfnamefont {Y.}~\bibnamefont {Paltiel}}, \bibinfo {author} {\bibfnamefont
  {W.}~\bibnamefont {Petuskey}}, \bibinfo {author} {\bibfnamefont {J.~C.}\ \bibnamefont {Ribeiro-Silva}}, \bibinfo {author} {\bibfnamefont {J.~J.}\ \bibnamefont {Saenz}}, \bibinfo {author} {\bibfnamefont {E.~J.~G.}\ \bibnamefont {Santos}}, \bibinfo {author} {\bibfnamefont {M.}~\bibnamefont {Solyanik-Gorgone}}, \bibinfo {author} {\bibfnamefont {V.~J.}\ \bibnamefont {Sorger}}, \bibinfo {author} {\bibfnamefont {D.~M.}\ \bibnamefont {Stemer}}, \bibinfo {author} {\bibfnamefont {J.~M.}\ \bibnamefont {Ugalde}}, \bibinfo {author} {\bibfnamefont {A.}~\bibnamefont {Valdes-Curiel}}, \bibinfo {author} {\bibfnamefont {S.}~\bibnamefont {Varela}}, \bibinfo {author} {\bibfnamefont {D.~H.}\ \bibnamefont {Waldeck}}, \bibinfo {author} {\bibfnamefont {M.~R.}\ \bibnamefont {Wasielewski}}, \bibinfo {author} {\bibfnamefont {P.~S.}\ \bibnamefont {Weiss}}, \bibinfo {author} {\bibfnamefont {H.}~\bibnamefont {Zacharias}},\ and\ \bibinfo {author} {\bibfnamefont {Q.~H.}\ \bibnamefont {Wang}},\ }\href@noop {} {\bibfield  {journal}
  {\bibinfo  {journal} {ACS Nano}\ }\textbf {\bibinfo {volume} {16}},\ \bibinfo {pages} {4989} (\bibinfo {year} {2022})}\BibitemShut {NoStop}%
\bibitem [{\citenamefont {Evers}\ \emph {et~al.}(2022)\citenamefont {Evers}, \citenamefont {Aharony}, \citenamefont {Bar-Gill}, \citenamefont {Entin-Wohlman}, \citenamefont {Hedegård}, \citenamefont {Hod}, \citenamefont {Jelinek}, \citenamefont {Kamieniarz}, \citenamefont {Lemeshko}, \citenamefont {Michaeli}, \citenamefont {Mujica}, \citenamefont {Naaman}, \citenamefont {Paltiel}, \citenamefont {Refaely-Abramson}, \citenamefont {Tal}, \citenamefont {Thijssen}, \citenamefont {Thoss}, \citenamefont {van Ruitenbeek}, \citenamefont {Venkataraman}, \citenamefont {Waldeck}, \citenamefont {Yan},\ and\ \citenamefont {Kronik}}]{Evers2022}%
  \BibitemOpen
  \bibfield  {author} {\bibinfo {author} {\bibfnamefont {F.}~\bibnamefont {Evers}}, \bibinfo {author} {\bibfnamefont {A.}~\bibnamefont {Aharony}}, \bibinfo {author} {\bibfnamefont {N.}~\bibnamefont {Bar-Gill}}, \bibinfo {author} {\bibfnamefont {O.}~\bibnamefont {Entin-Wohlman}}, \bibinfo {author} {\bibfnamefont {P.}~\bibnamefont {Hedegård}}, \bibinfo {author} {\bibfnamefont {O.}~\bibnamefont {Hod}}, \bibinfo {author} {\bibfnamefont {P.}~\bibnamefont {Jelinek}}, \bibinfo {author} {\bibfnamefont {G.}~\bibnamefont {Kamieniarz}}, \bibinfo {author} {\bibfnamefont {M.}~\bibnamefont {Lemeshko}}, \bibinfo {author} {\bibfnamefont {K.}~\bibnamefont {Michaeli}}, \bibinfo {author} {\bibfnamefont {V.}~\bibnamefont {Mujica}}, \bibinfo {author} {\bibfnamefont {R.}~\bibnamefont {Naaman}}, \bibinfo {author} {\bibfnamefont {Y.}~\bibnamefont {Paltiel}}, \bibinfo {author} {\bibfnamefont {S.}~\bibnamefont {Refaely-Abramson}}, \bibinfo {author} {\bibfnamefont {O.}~\bibnamefont {Tal}}, \bibinfo {author} {\bibfnamefont
  {J.}~\bibnamefont {Thijssen}}, \bibinfo {author} {\bibfnamefont {M.}~\bibnamefont {Thoss}}, \bibinfo {author} {\bibfnamefont {J.~M.}\ \bibnamefont {van Ruitenbeek}}, \bibinfo {author} {\bibfnamefont {L.}~\bibnamefont {Venkataraman}}, \bibinfo {author} {\bibfnamefont {D.~H.}\ \bibnamefont {Waldeck}}, \bibinfo {author} {\bibfnamefont {B.}~\bibnamefont {Yan}},\ and\ \bibinfo {author} {\bibfnamefont {L.}~\bibnamefont {Kronik}},\ }\href@noop {} {\bibfield  {journal} {\bibinfo  {journal} {Advanced Materials}\ }\textbf {\bibinfo {volume} {34}} (\bibinfo {year} {2022})}\BibitemShut {NoStop}%
\bibitem [{\citenamefont {Al-Bustami}\ \emph {et~al.}(2022)\citenamefont {Al-Bustami}, \citenamefont {Khaldi}, \citenamefont {Shoseyov}, \citenamefont {Yochelis}, \citenamefont {Killi}, \citenamefont {Berg}, \citenamefont {Gross}, \citenamefont {Paltiel},\ and\ \citenamefont {Yerushalmi}}]{Al-Bustami2022}%
  \BibitemOpen
  \bibfield  {author} {\bibinfo {author} {\bibfnamefont {H.}~\bibnamefont {Al-Bustami}}, \bibinfo {author} {\bibfnamefont {S.}~\bibnamefont {Khaldi}}, \bibinfo {author} {\bibfnamefont {O.}~\bibnamefont {Shoseyov}}, \bibinfo {author} {\bibfnamefont {S.}~\bibnamefont {Yochelis}}, \bibinfo {author} {\bibfnamefont {K.}~\bibnamefont {Killi}}, \bibinfo {author} {\bibfnamefont {I.}~\bibnamefont {Berg}}, \bibinfo {author} {\bibfnamefont {E.}~\bibnamefont {Gross}}, \bibinfo {author} {\bibfnamefont {Y.}~\bibnamefont {Paltiel}},\ and\ \bibinfo {author} {\bibfnamefont {R.}~\bibnamefont {Yerushalmi}},\ }\href@noop {} {\bibfield  {journal} {\bibinfo  {journal} {Nano Letters}\ }\textbf {\bibinfo {volume} {22}},\ \bibinfo {pages} {5022} (\bibinfo {year} {2022})}\BibitemShut {NoStop}%
\bibitem [{\citenamefont {Theiler}\ \emph {et~al.}(2023)\citenamefont {Theiler}, \citenamefont {Ritz}, \citenamefont {Hofmann},\ and\ \citenamefont {Stemmer}}]{Theiler2023}%
  \BibitemOpen
  \bibfield  {author} {\bibinfo {author} {\bibfnamefont {P.~M.}\ \bibnamefont {Theiler}}, \bibinfo {author} {\bibfnamefont {C.}~\bibnamefont {Ritz}}, \bibinfo {author} {\bibfnamefont {R.}~\bibnamefont {Hofmann}},\ and\ \bibinfo {author} {\bibfnamefont {A.}~\bibnamefont {Stemmer}},\ }\href@noop {} {\bibfield  {journal} {\bibinfo  {journal} {Nano Letters}\ }\textbf {\bibinfo {volume} {23}},\ \bibinfo {pages} {8280} (\bibinfo {year} {2023})}\BibitemShut {NoStop}%
\bibitem [{\citenamefont {Moharana}\ \emph {et~al.}(2025)\citenamefont {Moharana}, \citenamefont {Kapon}, \citenamefont {Kammerbauer}, \citenamefont {Anthofer}, \citenamefont {Yochelis}, \citenamefont {Shema}, \citenamefont {Gross}, \citenamefont {Kläui}, \citenamefont {Paltiel},\ and\ \citenamefont {Wittmann}}]{Moharana2025}%
  \BibitemOpen
  \bibfield  {author} {\bibinfo {author} {\bibfnamefont {A.}~\bibnamefont {Moharana}}, \bibinfo {author} {\bibfnamefont {Y.}~\bibnamefont {Kapon}}, \bibinfo {author} {\bibfnamefont {F.}~\bibnamefont {Kammerbauer}}, \bibinfo {author} {\bibfnamefont {D.}~\bibnamefont {Anthofer}}, \bibinfo {author} {\bibfnamefont {S.}~\bibnamefont {Yochelis}}, \bibinfo {author} {\bibfnamefont {H.}~\bibnamefont {Shema}}, \bibinfo {author} {\bibfnamefont {E.}~\bibnamefont {Gross}}, \bibinfo {author} {\bibfnamefont {M.}~\bibnamefont {Kläui}}, \bibinfo {author} {\bibfnamefont {Y.}~\bibnamefont {Paltiel}},\ and\ \bibinfo {author} {\bibfnamefont {A.}~\bibnamefont {Wittmann}},\ }\href {https://www.science.org} {\bibfield  {journal} {\bibinfo  {journal} {Sci. Adv}\ }\textbf {\bibinfo {volume} {11}},\ \bibinfo {pages} {4285} (\bibinfo {year} {2025})}\BibitemShut {NoStop}%
\bibitem [{\citenamefont {Inui}\ \emph {et~al.}(2020)\citenamefont {Inui}, \citenamefont {Aoki}, \citenamefont {Nishiue}, \citenamefont {Shiota}, \citenamefont {Kousaka}, \citenamefont {Shishido}, \citenamefont {Hirobe}, \citenamefont {Suda}, \citenamefont {Ohe}, \citenamefont {Kishine}, \citenamefont {Yamamoto},\ and\ \citenamefont {Togawa}}]{Inui2020}%
  \BibitemOpen
  \bibfield  {author} {\bibinfo {author} {\bibfnamefont {A.}~\bibnamefont {Inui}}, \bibinfo {author} {\bibfnamefont {R.}~\bibnamefont {Aoki}}, \bibinfo {author} {\bibfnamefont {Y.}~\bibnamefont {Nishiue}}, \bibinfo {author} {\bibfnamefont {K.}~\bibnamefont {Shiota}}, \bibinfo {author} {\bibfnamefont {Y.}~\bibnamefont {Kousaka}}, \bibinfo {author} {\bibfnamefont {H.}~\bibnamefont {Shishido}}, \bibinfo {author} {\bibfnamefont {D.}~\bibnamefont {Hirobe}}, \bibinfo {author} {\bibfnamefont {M.}~\bibnamefont {Suda}}, \bibinfo {author} {\bibfnamefont {J.-i.}\ \bibnamefont {Ohe}}, \bibinfo {author} {\bibfnamefont {J.-i.}\ \bibnamefont {Kishine}}, \bibinfo {author} {\bibfnamefont {H.~M.}\ \bibnamefont {Yamamoto}},\ and\ \bibinfo {author} {\bibfnamefont {Y.}~\bibnamefont {Togawa}},\ }\href@noop {} {\bibfield  {journal} {\bibinfo  {journal} {Physical Review Letters}\ }\textbf {\bibinfo {volume} {124}},\ \bibinfo {pages} {166602} (\bibinfo {year} {2020})}\BibitemShut {NoStop}%
\bibitem [{\citenamefont {Sun}\ \emph {et~al.}(2024{\natexlab{a}})\citenamefont {Sun}, \citenamefont {Park}, \citenamefont {Comstock}, \citenamefont {McConnell}, \citenamefont {Chen}, \citenamefont {Zhang}, \citenamefont {Beratan}, \citenamefont {You}, \citenamefont {Hoffmann}, \citenamefont {Yu}, \citenamefont {Diao},\ and\ \citenamefont {Sun}}]{Sun2024}%
  \BibitemOpen
  \bibfield  {author} {\bibinfo {author} {\bibfnamefont {R.}~\bibnamefont {Sun}}, \bibinfo {author} {\bibfnamefont {K.~S.}\ \bibnamefont {Park}}, \bibinfo {author} {\bibfnamefont {A.~H.}\ \bibnamefont {Comstock}}, \bibinfo {author} {\bibfnamefont {A.}~\bibnamefont {McConnell}}, \bibinfo {author} {\bibfnamefont {Y.~C.}\ \bibnamefont {Chen}}, \bibinfo {author} {\bibfnamefont {P.}~\bibnamefont {Zhang}}, \bibinfo {author} {\bibfnamefont {D.}~\bibnamefont {Beratan}}, \bibinfo {author} {\bibfnamefont {W.}~\bibnamefont {You}}, \bibinfo {author} {\bibfnamefont {A.}~\bibnamefont {Hoffmann}}, \bibinfo {author} {\bibfnamefont {Z.~G.}\ \bibnamefont {Yu}}, \bibinfo {author} {\bibfnamefont {Y.}~\bibnamefont {Diao}},\ and\ \bibinfo {author} {\bibfnamefont {D.}~\bibnamefont {Sun}},\ }\href@noop {} {\bibfield  {journal} {\bibinfo  {journal} {Nature Materials}\ }\textbf {\bibinfo {volume} {23}},\ \bibinfo {pages} {782} (\bibinfo {year} {2024}{\natexlab{a}})}\BibitemShut {NoStop}%
\bibitem [{\citenamefont {Abendroth}\ \emph {et~al.}(2019)\citenamefont {Abendroth}, \citenamefont {Cheung}, \citenamefont {Stemer}, \citenamefont {Hadri}, \citenamefont {Zhao}, \citenamefont {Fullerton},\ and\ \citenamefont {Weiss}}]{Abendroth2019}%
  \BibitemOpen
  \bibfield  {author} {\bibinfo {author} {\bibfnamefont {J.~M.}\ \bibnamefont {Abendroth}}, \bibinfo {author} {\bibfnamefont {K.~M.}\ \bibnamefont {Cheung}}, \bibinfo {author} {\bibfnamefont {D.~M.}\ \bibnamefont {Stemer}}, \bibinfo {author} {\bibfnamefont {M.~S.~E.}\ \bibnamefont {Hadri}}, \bibinfo {author} {\bibfnamefont {C.}~\bibnamefont {Zhao}}, \bibinfo {author} {\bibfnamefont {E.~E.}\ \bibnamefont {Fullerton}},\ and\ \bibinfo {author} {\bibfnamefont {P.~S.}\ \bibnamefont {Weiss}},\ }\href@noop {} {\bibfield  {journal} {\bibinfo  {journal} {Journal of the American Chemical Society}\ }\textbf {\bibinfo {volume} {141}},\ \bibinfo {pages} {38863} (\bibinfo {year} {2019})}\BibitemShut {NoStop}%
\bibitem [{\citenamefont {Liang}\ \emph {et~al.}(2022{\natexlab{a}})\citenamefont {Liang}, \citenamefont {Lihter},\ and\ \citenamefont {Lingenfelder}}]{Liang2022}%
  \BibitemOpen
  \bibfield  {author} {\bibinfo {author} {\bibfnamefont {Y.}~\bibnamefont {Liang}}, \bibinfo {author} {\bibfnamefont {M.}~\bibnamefont {Lihter}},\ and\ \bibinfo {author} {\bibfnamefont {M.}~\bibnamefont {Lingenfelder}},\ }\href@noop {} {\bibfield  {journal} {\bibinfo  {journal} {Israel Journal of Chemistry}\ }\textbf {\bibinfo {volume} {62}},\ \bibinfo {pages} {1} (\bibinfo {year} {2022}{\natexlab{a}})}\BibitemShut {NoStop}%
\bibitem [{\citenamefont {Bhowmick}\ \emph {et~al.}(2022)\citenamefont {Bhowmick}, \citenamefont {Das}, \citenamefont {Santra}, \citenamefont {Mondal}, \citenamefont {Tassinari}, \citenamefont {Schwarz}, \citenamefont {Diesendruck},\ and\ \citenamefont {Naaman}}]{Bhowmick2022}%
  \BibitemOpen
  \bibfield  {author} {\bibinfo {author} {\bibfnamefont {D.~K.}\ \bibnamefont {Bhowmick}}, \bibinfo {author} {\bibfnamefont {T.~K.}\ \bibnamefont {Das}}, \bibinfo {author} {\bibfnamefont {K.}~\bibnamefont {Santra}}, \bibinfo {author} {\bibfnamefont {A.~K.}\ \bibnamefont {Mondal}}, \bibinfo {author} {\bibfnamefont {F.}~\bibnamefont {Tassinari}}, \bibinfo {author} {\bibfnamefont {R.}~\bibnamefont {Schwarz}}, \bibinfo {author} {\bibfnamefont {C.~E.}\ \bibnamefont {Diesendruck}},\ and\ \bibinfo {author} {\bibfnamefont {R.}~\bibnamefont {Naaman}},\ }\href@noop {} {\bibfield  {journal} {\bibinfo  {journal} {Sci. Adv}\ }\textbf {\bibinfo {volume} {8}},\ \bibinfo {pages} {2727} (\bibinfo {year} {2022})}\BibitemShut {NoStop}%
\bibitem [{\citenamefont {{Ben Dor}}\ \emph {et~al.}(2017)\citenamefont {{Ben Dor}}, \citenamefont {Yochelis}, \citenamefont {Radko}, \citenamefont {Vankayala}, \citenamefont {Capua}, \citenamefont {Capua}, \citenamefont {Yang}, \citenamefont {Baczewski}, \citenamefont {Parkin}, \citenamefont {Naaman},\ and\ \citenamefont {Paltiel}}]{BenDor2017}%
  \BibitemOpen
  \bibfield  {author} {\bibinfo {author} {\bibfnamefont {O.}~\bibnamefont {{Ben Dor}}}, \bibinfo {author} {\bibfnamefont {S.}~\bibnamefont {Yochelis}}, \bibinfo {author} {\bibfnamefont {A.}~\bibnamefont {Radko}}, \bibinfo {author} {\bibfnamefont {K.}~\bibnamefont {Vankayala}}, \bibinfo {author} {\bibfnamefont {E.}~\bibnamefont {Capua}}, \bibinfo {author} {\bibfnamefont {A.}~\bibnamefont {Capua}}, \bibinfo {author} {\bibfnamefont {S.~H.}\ \bibnamefont {Yang}}, \bibinfo {author} {\bibfnamefont {L.~T.}\ \bibnamefont {Baczewski}}, \bibinfo {author} {\bibfnamefont {S.~S.~P.}\ \bibnamefont {Parkin}}, \bibinfo {author} {\bibfnamefont {R.}~\bibnamefont {Naaman}},\ and\ \bibinfo {author} {\bibfnamefont {Y.}~\bibnamefont {Paltiel}},\ }\href@noop {} {\bibfield  {journal} {\bibinfo  {journal} {Nature Communications}\ }\textbf {\bibinfo {volume} {8}},\ \bibinfo {pages} {1} (\bibinfo {year} {2017})}\BibitemShut {NoStop}%
\bibitem [{\citenamefont {Banerjee-Ghosh}\ \emph {et~al.}(2018)\citenamefont {Banerjee-Ghosh}, \citenamefont {Dor}, \citenamefont {Tassinari}, \citenamefont {Capua}, \citenamefont {Yochelis}, \citenamefont {Capua}, \citenamefont {Yang}, \citenamefont {Parkin}, \citenamefont {Sarkar}, \citenamefont {Kronik}, \citenamefont {Baczewski}, \citenamefont {Naaman},\ and\ \citenamefont {Paltiel}}]{Koyel2018}%
  \BibitemOpen
  \bibfield  {author} {\bibinfo {author} {\bibfnamefont {K.}~\bibnamefont {Banerjee-Ghosh}}, \bibinfo {author} {\bibfnamefont {O.~B.}\ \bibnamefont {Dor}}, \bibinfo {author} {\bibfnamefont {F.}~\bibnamefont {Tassinari}}, \bibinfo {author} {\bibfnamefont {E.}~\bibnamefont {Capua}}, \bibinfo {author} {\bibfnamefont {S.}~\bibnamefont {Yochelis}}, \bibinfo {author} {\bibfnamefont {A.}~\bibnamefont {Capua}}, \bibinfo {author} {\bibfnamefont {S.-H.}\ \bibnamefont {Yang}}, \bibinfo {author} {\bibfnamefont {S.~S.~P.}\ \bibnamefont {Parkin}}, \bibinfo {author} {\bibfnamefont {S.}~\bibnamefont {Sarkar}}, \bibinfo {author} {\bibfnamefont {L.}~\bibnamefont {Kronik}}, \bibinfo {author} {\bibfnamefont {L.~T.}\ \bibnamefont {Baczewski}}, \bibinfo {author} {\bibfnamefont {R.}~\bibnamefont {Naaman}},\ and\ \bibinfo {author} {\bibfnamefont {Y.}~\bibnamefont {Paltiel}},\ }\href@noop {} {\bibfield  {journal} {\bibinfo  {journal} {Science}\ }\textbf {\bibinfo {volume} {360}},\ \bibinfo {pages} {1331} (\bibinfo {year}
  {2018})}\BibitemShut {NoStop}%
\bibitem [{\citenamefont {Tirion}\ and\ \citenamefont {van Wees}(2024)}]{Tirion2024}%
  \BibitemOpen
  \bibfield  {author} {\bibinfo {author} {\bibfnamefont {S.~H.}\ \bibnamefont {Tirion}}\ and\ \bibinfo {author} {\bibfnamefont {B.~J.}\ \bibnamefont {van Wees}},\ }\href@noop {} {\bibfield  {journal} {\bibinfo  {journal} {ACS Nano}\ }\textbf {\bibinfo {volume} {18}},\ \bibinfo {pages} {6028} (\bibinfo {year} {2024})}\BibitemShut {NoStop}%
\bibitem [{\citenamefont {Xiong}(2018)}]{Xiong2018}%
  \BibitemOpen
  \bibfield  {author} {\bibinfo {author} {\bibfnamefont {Y.}~\bibnamefont {Xiong}},\ }\href@noop {} {\bibfield  {journal} {\bibinfo  {journal} {Journal of Physics Communications}\ }\textbf {\bibinfo {volume} {2}} (\bibinfo {year} {2018})}\BibitemShut {NoStop}%
\bibitem [{\citenamefont {Kunst}\ \emph {et~al.}(2018)\citenamefont {Kunst}, \citenamefont {Edvardsson}, \citenamefont {Budich},\ and\ \citenamefont {Bergholtz}}]{Kunst2018}%
  \BibitemOpen
  \bibfield  {author} {\bibinfo {author} {\bibfnamefont {F.~K.}\ \bibnamefont {Kunst}}, \bibinfo {author} {\bibfnamefont {E.}~\bibnamefont {Edvardsson}}, \bibinfo {author} {\bibfnamefont {J.~C.}\ \bibnamefont {Budich}},\ and\ \bibinfo {author} {\bibfnamefont {E.~J.}\ \bibnamefont {Bergholtz}},\ }\href@noop {} {\bibfield  {journal} {\bibinfo  {journal} {Physical Review Letters}\ }\textbf {\bibinfo {volume} {121}} (\bibinfo {year} {2018})}\BibitemShut {NoStop}%
\bibitem [{\citenamefont {Streater}\ and\ \citenamefont {Wigthman}(1964)}]{Streater1964}%
  \BibitemOpen
  \bibfield  {author} {\bibinfo {author} {\bibfnamefont {R.}~\bibnamefont {Streater}}\ and\ \bibinfo {author} {\bibfnamefont {A.}~\bibnamefont {Wigthman}},\ }\href@noop {} {\emph {\bibinfo {title} {{PCT, spin and statistics, and all that}}}}\ (\bibinfo  {publisher} {Princeton University Press},\ \bibinfo {address} {Princeton},\ \bibinfo {year} {1964})\ p.\ \bibinfo {pages} {218}\BibitemShut {NoStop}%
\bibitem [{\citenamefont {Dresselhaus}\ \emph {et~al.}(2008)\citenamefont {Dresselhaus}, \citenamefont {Dresselhaus},\ and\ \citenamefont {Jorio}}]{Dresselhaus2008}%
  \BibitemOpen
  \bibfield  {author} {\bibinfo {author} {\bibfnamefont {M.~S.}\ \bibnamefont {Dresselhaus}}, \bibinfo {author} {\bibfnamefont {G.}~\bibnamefont {Dresselhaus}},\ and\ \bibinfo {author} {\bibfnamefont {A.}~\bibnamefont {Jorio}},\ }\href@noop {} {\emph {\bibinfo {title} {Group theory : application to the physics of condensed matter}}}\ (\bibinfo  {publisher} {Springer-Verlag},\ \bibinfo {year} {2008})\ p.\ \bibinfo {pages} {582}\BibitemShut {NoStop}%
\bibitem [{\citenamefont {Liu}\ \emph {et~al.}(2022)\citenamefont {Liu}, \citenamefont {Li}, \citenamefont {Han}, \citenamefont {Wan},\ and\ \citenamefont {Liu}}]{Liu2022}%
  \BibitemOpen
  \bibfield  {author} {\bibinfo {author} {\bibfnamefont {P.}~\bibnamefont {Liu}}, \bibinfo {author} {\bibfnamefont {J.}~\bibnamefont {Li}}, \bibinfo {author} {\bibfnamefont {J.}~\bibnamefont {Han}}, \bibinfo {author} {\bibfnamefont {X.}~\bibnamefont {Wan}},\ and\ \bibinfo {author} {\bibfnamefont {Q.}~\bibnamefont {Liu}},\ }\href@noop {} {\bibfield  {journal} {\bibinfo  {journal} {Physical Review X}\ }\textbf {\bibinfo {volume} {12}} (\bibinfo {year} {2022})}\BibitemShut {NoStop}%
\bibitem [{\citenamefont {Capozziello}\ and\ \citenamefont {Lattanzi}(2003)}]{Capozziello2003}%
  \BibitemOpen
  \bibfield  {author} {\bibinfo {author} {\bibfnamefont {S.}~\bibnamefont {Capozziello}}\ and\ \bibinfo {author} {\bibfnamefont {A.}~\bibnamefont {Lattanzi}},\ }\href@noop {} {\bibfield  {journal} {\bibinfo  {journal} {Chirality}\ }\textbf {\bibinfo {volume} {15}},\ \bibinfo {pages} {466} (\bibinfo {year} {2003})}\BibitemShut {NoStop}%
\bibitem [{\citenamefont {Kusunose}\ \emph {et~al.}(2024)\citenamefont {Kusunose}, \citenamefont {Kishine},\ and\ \citenamefont {Yamamoto}}]{Kusunose2024}%
  \BibitemOpen
  \bibfield  {author} {\bibinfo {author} {\bibfnamefont {H.}~\bibnamefont {Kusunose}}, \bibinfo {author} {\bibfnamefont {J.~I.}\ \bibnamefont {Kishine}},\ and\ \bibinfo {author} {\bibfnamefont {H.~M.}\ \bibnamefont {Yamamoto}},\ }\href@noop {} {\bibfield  {journal} {\bibinfo  {journal} {Applied Physics Letters}\ }\textbf {\bibinfo {volume} {124}} (\bibinfo {year} {2024})}\BibitemShut {NoStop}%
\bibitem [{\citenamefont {Wang}\ and\ \citenamefont {Hazzard}(2025)}]{Wang2025}%
  \BibitemOpen
  \bibfield  {author} {\bibinfo {author} {\bibfnamefont {Z.}~\bibnamefont {Wang}}\ and\ \bibinfo {author} {\bibfnamefont {K.~R.~A.}\ \bibnamefont {Hazzard}},\ }\href@noop {} {\bibfield  {journal} {\bibinfo  {journal} {Nature}\ }\textbf {\bibinfo {volume} {637}},\ \bibinfo {pages} {314} (\bibinfo {year} {2025})}\BibitemShut {NoStop}%
\bibitem [{\citenamefont {Hoffmann}\ and\ \citenamefont {Blügel}(2020)}]{Hoffmann2020}%
  \BibitemOpen
  \bibfield  {author} {\bibinfo {author} {\bibfnamefont {M.}~\bibnamefont {Hoffmann}}\ and\ \bibinfo {author} {\bibfnamefont {S.}~\bibnamefont {Blügel}},\ }\href@noop {} {\bibfield  {journal} {\bibinfo  {journal} {Physical Review B}\ }\textbf {\bibinfo {volume} {101}} (\bibinfo {year} {2020})}\BibitemShut {NoStop}%
\bibitem [{\citenamefont {Dirac}(1928)}]{Dirac1928}%
  \BibitemOpen
  \bibfield  {author} {\bibinfo {author} {\bibfnamefont {P.~A.~M.}\ \bibnamefont {Dirac}},\ }\href@noop {} {\bibfield  {journal} {\bibinfo  {journal} {Roy. Soc. Proc.,' A}\ }\textbf {\bibinfo {volume} {117}},\ \bibinfo {pages} {610} (\bibinfo {year} {1928})}\BibitemShut {NoStop}%
\bibitem [{\citenamefont {Watson}\ and\ \citenamefont {Musielak}(2021)}]{Watson2021}%
  \BibitemOpen
  \bibfield  {author} {\bibinfo {author} {\bibfnamefont {T.~B.}\ \bibnamefont {Watson}}\ and\ \bibinfo {author} {\bibfnamefont {Z.~E.}\ \bibnamefont {Musielak}},\ }\href@noop {} {\bibfield  {journal} {\bibinfo  {journal} {Symmetry}\ }\textbf {\bibinfo {volume} {13}} (\bibinfo {year} {2021})}\BibitemShut {NoStop}%
\bibitem [{\citenamefont {Barron}(2020)}]{Barron2020}%
  \BibitemOpen
  \bibfield  {author} {\bibinfo {author} {\bibfnamefont {L.~D.}\ \bibnamefont {Barron}},\ }\href@noop {} {\bibfield  {journal} {\bibinfo  {journal} {Magnetochemistry}\ }\textbf {\bibinfo {volume} {6}},\ \bibinfo {pages} {1} (\bibinfo {year} {2020})}\BibitemShut {NoStop}%
\bibitem [{\citenamefont {Hatano}\ and\ \citenamefont {Nelson}(1996)}]{Hatano1996}%
  \BibitemOpen
  \bibfield  {author} {\bibinfo {author} {\bibfnamefont {N.}~\bibnamefont {Hatano}}\ and\ \bibinfo {author} {\bibfnamefont {D.~R.}\ \bibnamefont {Nelson}},\ }\href@noop {} {\bibfield  {journal} {\bibinfo  {journal} {Physcial Review Letters}\ }\textbf {\bibinfo {volume} {77}},\ \bibinfo {pages} {570} (\bibinfo {year} {1996})}\BibitemShut {NoStop}%
\bibitem [{\citenamefont {Bender}(2007)}]{Bender2007}%
  \BibitemOpen
  \bibfield  {author} {\bibinfo {author} {\bibfnamefont {C.~M.}\ \bibnamefont {Bender}},\ }\href@noop {} {\bibfield  {journal} {\bibinfo  {journal} {Reports on Progress in Physics}\ }\textbf {\bibinfo {volume} {70}},\ \bibinfo {pages} {947} (\bibinfo {year} {2007})}\BibitemShut {NoStop}%
\bibitem [{\citenamefont {Ashida}\ \emph {et~al.}(2020)\citenamefont {Ashida}, \citenamefont {Gong},\ and\ \citenamefont {Ueda}}]{Ashida2020}%
  \BibitemOpen
  \bibfield  {author} {\bibinfo {author} {\bibfnamefont {Y.}~\bibnamefont {Ashida}}, \bibinfo {author} {\bibfnamefont {Z.}~\bibnamefont {Gong}},\ and\ \bibinfo {author} {\bibfnamefont {M.}~\bibnamefont {Ueda}},\ }\href@noop {} {\bibfield  {journal} {\bibinfo  {journal} {Advances in Physics}\ }\textbf {\bibinfo {volume} {69}} (\bibinfo {year} {2020})}\BibitemShut {NoStop}%
\bibitem [{\citenamefont {Smolinsky}\ \emph {et~al.}(2019)\citenamefont {Smolinsky}, \citenamefont {Neubauer}, \citenamefont {Kumar}, \citenamefont {Yochelis}, \citenamefont {Capua}, \citenamefont {Carmieli}, \citenamefont {Paltiel}, \citenamefont {Naaman},\ and\ \citenamefont {Michaeli}}]{Smolinsky2019}%
  \BibitemOpen
  \bibfield  {author} {\bibinfo {author} {\bibfnamefont {E.~Z.}\ \bibnamefont {Smolinsky}}, \bibinfo {author} {\bibfnamefont {A.}~\bibnamefont {Neubauer}}, \bibinfo {author} {\bibfnamefont {A.}~\bibnamefont {Kumar}}, \bibinfo {author} {\bibfnamefont {S.}~\bibnamefont {Yochelis}}, \bibinfo {author} {\bibfnamefont {E.}~\bibnamefont {Capua}}, \bibinfo {author} {\bibfnamefont {R.}~\bibnamefont {Carmieli}}, \bibinfo {author} {\bibfnamefont {Y.}~\bibnamefont {Paltiel}}, \bibinfo {author} {\bibfnamefont {R.}~\bibnamefont {Naaman}},\ and\ \bibinfo {author} {\bibfnamefont {K.}~\bibnamefont {Michaeli}},\ }\href@noop {} {\bibfield  {journal} {\bibinfo  {journal} {Journal of Physical Chemistry Letters}\ }\textbf {\bibinfo {volume} {10}},\ \bibinfo {pages} {1139} (\bibinfo {year} {2019})}\BibitemShut {NoStop}%
\bibitem [{\citenamefont {Koplovitz}\ \emph {et~al.}(2019)\citenamefont {Koplovitz}, \citenamefont {Leitus}, \citenamefont {Ghosh}, \citenamefont {Bloom}, \citenamefont {Yochelis}, \citenamefont {Rotem}, \citenamefont {Vischio}, \citenamefont {Striccoli}, \citenamefont {Fanizza}, \citenamefont {Naaman}, \citenamefont {Waldeck}, \citenamefont {Porath},\ and\ \citenamefont {Paltiel}}]{Koplovitz2019}%
  \BibitemOpen
  \bibfield  {author} {\bibinfo {author} {\bibfnamefont {G.}~\bibnamefont {Koplovitz}}, \bibinfo {author} {\bibfnamefont {G.}~\bibnamefont {Leitus}}, \bibinfo {author} {\bibfnamefont {S.}~\bibnamefont {Ghosh}}, \bibinfo {author} {\bibfnamefont {B.~P.}\ \bibnamefont {Bloom}}, \bibinfo {author} {\bibfnamefont {S.}~\bibnamefont {Yochelis}}, \bibinfo {author} {\bibfnamefont {D.}~\bibnamefont {Rotem}}, \bibinfo {author} {\bibfnamefont {F.}~\bibnamefont {Vischio}}, \bibinfo {author} {\bibfnamefont {M.}~\bibnamefont {Striccoli}}, \bibinfo {author} {\bibfnamefont {E.}~\bibnamefont {Fanizza}}, \bibinfo {author} {\bibfnamefont {R.}~\bibnamefont {Naaman}}, \bibinfo {author} {\bibfnamefont {D.~H.}\ \bibnamefont {Waldeck}}, \bibinfo {author} {\bibfnamefont {D.}~\bibnamefont {Porath}},\ and\ \bibinfo {author} {\bibfnamefont {Y.}~\bibnamefont {Paltiel}},\ }\href@noop {} {\bibfield  {journal} {\bibinfo  {journal} {Small}\ }\textbf {\bibinfo {volume} {15}} (\bibinfo {year} {2019})}\BibitemShut {NoStop}%
\bibitem [{\citenamefont {Meirzada}\ \emph {et~al.}(2021)\citenamefont {Meirzada}, \citenamefont {Sukenik}, \citenamefont {Haim}, \citenamefont {Yochelis}, \citenamefont {Baczewski}, \citenamefont {Paltiel},\ and\ \citenamefont {Bar-Gill}}]{Meirzada2021}%
  \BibitemOpen
  \bibfield  {author} {\bibinfo {author} {\bibfnamefont {I.}~\bibnamefont {Meirzada}}, \bibinfo {author} {\bibfnamefont {N.}~\bibnamefont {Sukenik}}, \bibinfo {author} {\bibfnamefont {G.}~\bibnamefont {Haim}}, \bibinfo {author} {\bibfnamefont {S.}~\bibnamefont {Yochelis}}, \bibinfo {author} {\bibfnamefont {L.~T.}\ \bibnamefont {Baczewski}}, \bibinfo {author} {\bibfnamefont {Y.}~\bibnamefont {Paltiel}},\ and\ \bibinfo {author} {\bibfnamefont {N.}~\bibnamefont {Bar-Gill}},\ }\href@noop {} {\bibfield  {journal} {\bibinfo  {journal} {ACS Nano}\ }\textbf {\bibinfo {volume} {15}},\ \bibinfo {pages} {5574} (\bibinfo {year} {2021})}\BibitemShut {NoStop}%
\bibitem [{\citenamefont {Nakajima}\ \emph {et~al.}(2023)\citenamefont {Nakajima}, \citenamefont {Hirobe}, \citenamefont {Kawaguchi}, \citenamefont {Nabei}, \citenamefont {Sato}, \citenamefont {Narushima}, \citenamefont {Okamoto},\ and\ \citenamefont {Yamamoto}}]{Nakajima2023}%
  \BibitemOpen
  \bibfield  {author} {\bibinfo {author} {\bibfnamefont {R.}~\bibnamefont {Nakajima}}, \bibinfo {author} {\bibfnamefont {D.}~\bibnamefont {Hirobe}}, \bibinfo {author} {\bibfnamefont {G.}~\bibnamefont {Kawaguchi}}, \bibinfo {author} {\bibfnamefont {Y.}~\bibnamefont {Nabei}}, \bibinfo {author} {\bibfnamefont {T.}~\bibnamefont {Sato}}, \bibinfo {author} {\bibfnamefont {T.}~\bibnamefont {Narushima}}, \bibinfo {author} {\bibfnamefont {H.}~\bibnamefont {Okamoto}},\ and\ \bibinfo {author} {\bibfnamefont {H.~M.}\ \bibnamefont {Yamamoto}},\ }\href@noop {} {\bibfield  {journal} {\bibinfo  {journal} {Nature}\ }\textbf {\bibinfo {volume} {613}},\ \bibinfo {pages} {479} (\bibinfo {year} {2023})}\BibitemShut {NoStop}%
\bibitem [{\citenamefont {Ghosh}\ \emph {et~al.}(2020)\citenamefont {Ghosh}, \citenamefont {Mishra}, \citenamefont {Avigad}, \citenamefont {Bloom}, \citenamefont {Baczewski}, \citenamefont {Yochelis}, \citenamefont {Paltiel}, \citenamefont {Naaman},\ and\ \citenamefont {Waldeck}}]{Ghosh2020}%
  \BibitemOpen
  \bibfield  {author} {\bibinfo {author} {\bibfnamefont {S.}~\bibnamefont {Ghosh}}, \bibinfo {author} {\bibfnamefont {S.}~\bibnamefont {Mishra}}, \bibinfo {author} {\bibfnamefont {E.}~\bibnamefont {Avigad}}, \bibinfo {author} {\bibfnamefont {B.~P.}\ \bibnamefont {Bloom}}, \bibinfo {author} {\bibfnamefont {L.~T.}\ \bibnamefont {Baczewski}}, \bibinfo {author} {\bibfnamefont {S.}~\bibnamefont {Yochelis}}, \bibinfo {author} {\bibfnamefont {Y.}~\bibnamefont {Paltiel}}, \bibinfo {author} {\bibfnamefont {R.}~\bibnamefont {Naaman}},\ and\ \bibinfo {author} {\bibfnamefont {D.~H.}\ \bibnamefont {Waldeck}},\ }\href@noop {} {\bibfield  {journal} {\bibinfo  {journal} {Journal of Physical Chemistry Letters}\ }\textbf {\bibinfo {volume} {11}},\ \bibinfo {pages} {1550} (\bibinfo {year} {2020})}\BibitemShut {NoStop}%
\bibitem [{\citenamefont {Kapon}\ \emph {et~al.}(2021)\citenamefont {Kapon}, \citenamefont {Saha}, \citenamefont {Duanis-Assaf}, \citenamefont {Stuyver}, \citenamefont {Ziv}, \citenamefont {Metzger}, \citenamefont {Yochelis}, \citenamefont {Shaik}, \citenamefont {Naaman}, \citenamefont {Reches},\ and\ \citenamefont {Paltiel}}]{Kapon2021}%
  \BibitemOpen
  \bibfield  {author} {\bibinfo {author} {\bibfnamefont {Y.}~\bibnamefont {Kapon}}, \bibinfo {author} {\bibfnamefont {A.}~\bibnamefont {Saha}}, \bibinfo {author} {\bibfnamefont {T.}~\bibnamefont {Duanis-Assaf}}, \bibinfo {author} {\bibfnamefont {T.}~\bibnamefont {Stuyver}}, \bibinfo {author} {\bibfnamefont {A.}~\bibnamefont {Ziv}}, \bibinfo {author} {\bibfnamefont {T.}~\bibnamefont {Metzger}}, \bibinfo {author} {\bibfnamefont {S.}~\bibnamefont {Yochelis}}, \bibinfo {author} {\bibfnamefont {S.}~\bibnamefont {Shaik}}, \bibinfo {author} {\bibfnamefont {R.}~\bibnamefont {Naaman}}, \bibinfo {author} {\bibfnamefont {M.}~\bibnamefont {Reches}},\ and\ \bibinfo {author} {\bibfnamefont {Y.}~\bibnamefont {Paltiel}},\ }\href@noop {} {\bibfield  {journal} {\bibinfo  {journal} {Chem}\ }\textbf {\bibinfo {volume} {7}},\ \bibinfo {pages} {2787} (\bibinfo {year} {2021})}\BibitemShut {NoStop}%
\bibitem [{\citenamefont {Naaman}\ and\ \citenamefont {Waldeck}(2024)}]{Namaan2024}%
  \BibitemOpen
  \bibfield  {author} {\bibinfo {author} {\bibfnamefont {R.}~\bibnamefont {Naaman}}\ and\ \bibinfo {author} {\bibfnamefont {D.~H.}\ \bibnamefont {Waldeck}},\ }in\ \href {https://www.sciencedirect.com/science/article/pii/B9780323856690000106} {\emph {\bibinfo {booktitle} {Encyclopedia of Solid-Liquid Interfaces (First Edition)}}},\ \bibinfo {editor} {edited by\ \bibinfo {editor} {\bibfnamefont {K.}~\bibnamefont {Wandelt}}\ and\ \bibinfo {editor} {\bibfnamefont {G.}~\bibnamefont {Bussetti}}}\ (\bibinfo  {publisher} {Elsevier},\ \bibinfo {address} {Oxford},\ \bibinfo {year} {2024})\ \bibinfo {edition} {first edition}\ ed.,\ pp.\ \bibinfo {pages} {267--277}\BibitemShut {NoStop}%
\bibitem [{\citenamefont {Hautzinger}\ \emph {et~al.}(2024)\citenamefont {Hautzinger}, \citenamefont {Pan}, \citenamefont {Hayden}, \citenamefont {Ye}, \citenamefont {Jiang}, \citenamefont {Wilson}, \citenamefont {Phillips}, \citenamefont {Dong}, \citenamefont {Raulerson}, \citenamefont {Leahy}, \citenamefont {Jiang}, \citenamefont {Blackburn}, \citenamefont {Luther}, \citenamefont {Lu}, \citenamefont {Jungjohann}, \citenamefont {Vardeny}, \citenamefont {Berry}, \citenamefont {Alberi},\ and\ \citenamefont {Beard}}]{Hautzinger2024}%
  \BibitemOpen
  \bibfield  {author} {\bibinfo {author} {\bibfnamefont {M.~P.}\ \bibnamefont {Hautzinger}}, \bibinfo {author} {\bibfnamefont {X.}~\bibnamefont {Pan}}, \bibinfo {author} {\bibfnamefont {S.~C.}\ \bibnamefont {Hayden}}, \bibinfo {author} {\bibfnamefont {J.~Y.}\ \bibnamefont {Ye}}, \bibinfo {author} {\bibfnamefont {Q.}~\bibnamefont {Jiang}}, \bibinfo {author} {\bibfnamefont {M.~J.}\ \bibnamefont {Wilson}}, \bibinfo {author} {\bibfnamefont {A.~J.}\ \bibnamefont {Phillips}}, \bibinfo {author} {\bibfnamefont {Y.}~\bibnamefont {Dong}}, \bibinfo {author} {\bibfnamefont {E.~K.}\ \bibnamefont {Raulerson}}, \bibinfo {author} {\bibfnamefont {I.~A.}\ \bibnamefont {Leahy}}, \bibinfo {author} {\bibfnamefont {C.~S.}\ \bibnamefont {Jiang}}, \bibinfo {author} {\bibfnamefont {J.~L.}\ \bibnamefont {Blackburn}}, \bibinfo {author} {\bibfnamefont {J.~M.}\ \bibnamefont {Luther}}, \bibinfo {author} {\bibfnamefont {Y.}~\bibnamefont {Lu}}, \bibinfo {author} {\bibfnamefont {K.}~\bibnamefont {Jungjohann}}, \bibinfo {author}
  {\bibfnamefont {Z.~V.}\ \bibnamefont {Vardeny}}, \bibinfo {author} {\bibfnamefont {J.~J.}\ \bibnamefont {Berry}}, \bibinfo {author} {\bibfnamefont {K.}~\bibnamefont {Alberi}},\ and\ \bibinfo {author} {\bibfnamefont {M.~C.}\ \bibnamefont {Beard}},\ }\href@noop {} {\bibfield  {journal} {\bibinfo  {journal} {Nature}\ }\textbf {\bibinfo {volume} {631}},\ \bibinfo {pages} {307} (\bibinfo {year} {2024})}\BibitemShut {NoStop}%
\bibitem [{\citenamefont {Liang}\ \emph {et~al.}(2022{\natexlab{b}})\citenamefont {Liang}, \citenamefont {Banjac}, \citenamefont {Martin}, \citenamefont {Zigon}, \citenamefont {Lee}, \citenamefont {Vanthuyne}, \citenamefont {Garcés-Pineda}, \citenamefont {Galán-Mascarós}, \citenamefont {Hu}, \citenamefont {Avarvari},\ and\ \citenamefont {Lingenfelder}}]{Liang2022a}%
  \BibitemOpen
  \bibfield  {author} {\bibinfo {author} {\bibfnamefont {Y.}~\bibnamefont {Liang}}, \bibinfo {author} {\bibfnamefont {K.}~\bibnamefont {Banjac}}, \bibinfo {author} {\bibfnamefont {K.}~\bibnamefont {Martin}}, \bibinfo {author} {\bibfnamefont {N.}~\bibnamefont {Zigon}}, \bibinfo {author} {\bibfnamefont {S.}~\bibnamefont {Lee}}, \bibinfo {author} {\bibfnamefont {N.}~\bibnamefont {Vanthuyne}}, \bibinfo {author} {\bibfnamefont {F.~A.}\ \bibnamefont {Garcés-Pineda}}, \bibinfo {author} {\bibfnamefont {J.~R.}\ \bibnamefont {Galán-Mascarós}}, \bibinfo {author} {\bibfnamefont {X.}~\bibnamefont {Hu}}, \bibinfo {author} {\bibfnamefont {N.}~\bibnamefont {Avarvari}},\ and\ \bibinfo {author} {\bibfnamefont {M.}~\bibnamefont {Lingenfelder}},\ }\href@noop {} {\bibfield  {journal} {\bibinfo  {journal} {Nature Communications}\ }\textbf {\bibinfo {volume} {13}} (\bibinfo {year} {2022}{\natexlab{b}})}\BibitemShut {NoStop}%
\bibitem [{\citenamefont {Bardarson}(2008)}]{Bardarson2008}%
  \BibitemOpen
  \bibfield  {author} {\bibinfo {author} {\bibfnamefont {J.~H.}\ \bibnamefont {Bardarson}},\ }\href@noop {} {\bibfield  {journal} {\bibinfo  {journal} {Journal of Physics A: Mathematical and Theoretical}\ }\textbf {\bibinfo {volume} {41}} (\bibinfo {year} {2008})}\BibitemShut {NoStop}%
\bibitem [{\citenamefont {M{\"{o}}llers}\ \emph {et~al.}(2022)\citenamefont {M{\"{o}}llers}, \citenamefont {G{\"{o}}hler},\ and\ \citenamefont {Zacharias}}]{Mollers2022}%
  \BibitemOpen
  \bibfield  {author} {\bibinfo {author} {\bibfnamefont {P.~V.}\ \bibnamefont {M{\"{o}}llers}}, \bibinfo {author} {\bibfnamefont {B.}~\bibnamefont {G{\"{o}}hler}},\ and\ \bibinfo {author} {\bibfnamefont {H.}~\bibnamefont {Zacharias}},\ }\href@noop {} {\bibfield  {journal} {\bibinfo  {journal} {Israel Journal of Chemistry}\ }\textbf {\bibinfo {volume} {62}},\ \bibinfo {pages} {1} (\bibinfo {year} {2022})}\BibitemShut {NoStop}%
\bibitem [{\citenamefont {Eckvahl}\ \emph {et~al.}(2023)\citenamefont {Eckvahl}, \citenamefont {Tcyrulnikov}, \citenamefont {Chiesa}, \citenamefont {Bradley}, \citenamefont {Young}, \citenamefont {Carretta}, \citenamefont {Krzyaniak},\ and\ \citenamefont {Wasielewski}}]{Eckvahl2023}%
  \BibitemOpen
  \bibfield  {author} {\bibinfo {author} {\bibfnamefont {H.~J.}\ \bibnamefont {Eckvahl}}, \bibinfo {author} {\bibfnamefont {N.~A.}\ \bibnamefont {Tcyrulnikov}}, \bibinfo {author} {\bibfnamefont {A.}~\bibnamefont {Chiesa}}, \bibinfo {author} {\bibfnamefont {J.~M.}\ \bibnamefont {Bradley}}, \bibinfo {author} {\bibfnamefont {R.~M.}\ \bibnamefont {Young}}, \bibinfo {author} {\bibfnamefont {S.}~\bibnamefont {Carretta}}, \bibinfo {author} {\bibfnamefont {M.~D.}\ \bibnamefont {Krzyaniak}},\ and\ \bibinfo {author} {\bibfnamefont {M.~R.}\ \bibnamefont {Wasielewski}},\ }\href {https://www.science.org} {\bibfield  {journal} {\bibinfo  {journal} {Science}\ }\textbf {\bibinfo {volume} {382}},\ \bibinfo {pages} {2025} (\bibinfo {year} {2023})}\BibitemShut {NoStop}%
\bibitem [{\citenamefont {Sun}\ \emph {et~al.}(2024{\natexlab{b}})\citenamefont {Sun}, \citenamefont {Wang}, \citenamefont {Bloom}, \citenamefont {Comstock}, \citenamefont {Yang}, \citenamefont {McConnell}, \citenamefont {Clever}, \citenamefont {Molitoris}, \citenamefont {Lamont}, \citenamefont {Cheng}, \citenamefont {Yuan}, \citenamefont {Zhang}, \citenamefont {Hoffmann}, \citenamefont {Liu}, \citenamefont {Waldeck},\ and\ \citenamefont {Sun}}]{Sun2024_1}%
  \BibitemOpen
  \bibfield  {author} {\bibinfo {author} {\bibfnamefont {R.}~\bibnamefont {Sun}}, \bibinfo {author} {\bibfnamefont {Z.}~\bibnamefont {Wang}}, \bibinfo {author} {\bibfnamefont {B.~P.}\ \bibnamefont {Bloom}}, \bibinfo {author} {\bibfnamefont {A.~H.}\ \bibnamefont {Comstock}}, \bibinfo {author} {\bibfnamefont {C.}~\bibnamefont {Yang}}, \bibinfo {author} {\bibfnamefont {A.}~\bibnamefont {McConnell}}, \bibinfo {author} {\bibfnamefont {C.}~\bibnamefont {Clever}}, \bibinfo {author} {\bibfnamefont {M.}~\bibnamefont {Molitoris}}, \bibinfo {author} {\bibfnamefont {D.}~\bibnamefont {Lamont}}, \bibinfo {author} {\bibfnamefont {Z.~H.}\ \bibnamefont {Cheng}}, \bibinfo {author} {\bibfnamefont {Z.}~\bibnamefont {Yuan}}, \bibinfo {author} {\bibfnamefont {W.}~\bibnamefont {Zhang}}, \bibinfo {author} {\bibfnamefont {A.}~\bibnamefont {Hoffmann}}, \bibinfo {author} {\bibfnamefont {J.}~\bibnamefont {Liu}}, \bibinfo {author} {\bibfnamefont {D.~H.}\ \bibnamefont {Waldeck}},\ and\ \bibinfo {author} {\bibfnamefont {D.}~\bibnamefont
  {Sun}},\ }\href@noop {} {\bibfield  {journal} {\bibinfo  {journal} {Science Advances}\ }\textbf {\bibinfo {volume} {10}} (\bibinfo {year} {2024}{\natexlab{b}})}\BibitemShut {NoStop}%
\bibitem [{\citenamefont {Gross}\ and\ \citenamefont {Bloch}(2017)}]{Gross2017}%
  \BibitemOpen
  \bibfield  {author} {\bibinfo {author} {\bibfnamefont {C.}~\bibnamefont {Gross}}\ and\ \bibinfo {author} {\bibfnamefont {I.}~\bibnamefont {Bloch}},\ }\href {https://www.science.org} {\bibfield  {journal} {\bibinfo  {journal} {Science}\ ,\ \bibinfo {pages} {995}} (\bibinfo {year} {2017})}\BibitemShut {NoStop}%
\bibitem [{\citenamefont {Lifshitz}\ and\ \citenamefont {Pitaevskii}(1981)}]{Lifshitz1981}%
  \BibitemOpen
  \bibfield  {author} {\bibinfo {author} {\bibfnamefont {E.~M.}\ \bibnamefont {Lifshitz}}\ and\ \bibinfo {author} {\bibfnamefont {L.~P.}\ \bibnamefont {Pitaevskii}},\ }\href@noop {} {\emph {\bibinfo {title} {{Physical kinetics. Vol. 10 (1st ed.)}}}}\ (\bibinfo  {publisher} {Pergamon Press},\ \bibinfo {year} {1981})\ p.\ \bibinfo {pages} {462}\BibitemShut {NoStop}%
\end{thebibliography}
\end{document}